\documentclass[12pt,a4paper]{article}
\usepackage[english]{babel}
\usepackage[utf8]{inputenc}
\usepackage[dvips]{graphics}
\usepackage{longtable}
\usepackage{lscape}
\usepackage{footnote}
\makesavenoteenv{table}
\makesavenoteenv{tabular}
\makesavenoteenv{figure}

\catcode`\{=1
\catcode`\}=2
\def\tfig#1{figure~#1}
\def\ffig#1#2#3#4{\begin{figure}[#2]
\hfil \includegraphics{#3} \hfil \newline {\centerline {\it figure #1#4}}
\end{figure}}
\def\rfig#1#2#3#4{\tfig{#1}\ffig{#1}{#2}{#3}{#4}}

\def\lref#1{{\tt [#1]}}

\def\term#1{{\it #1}}

\def\cite#1{{\it ``#1''}}

\usepackage[OT2,OT1]{fontenc}
\newcommand\cyr
{
\renewcommand\rmdefault{wncyr}
\renewcommand\sfdefault{wncyss}
\renewcommand\encodingdefault{OT2}
\normalfont
\selectfont
}
\DeclareTextFontCommand{\textcyr}{\cyr}

\title{Parallel Architecture Hardware and \\
  General Purpose Operating System Co-design}
\author{Oskar Schirmer}
\date{Göttingen, 2018-07-10}
\begin{document}

\maketitle
\vfill

\section*{Abstract}

Because most optimisations to achieve higher
computational performance eventually are limited,
parallelism that scales is required.
Parallelised hardware alone is not sufficient,
but software that matches the architecture is required
to gain best performance.
For decades now, hardware design has been guided by
the basic design of existing software,
to avoid the higher cost to redesign the latter.
In doing so, however, quite a variety of superior concepts is excluded a priori.
Consequently, co-design of both hardware and software
is crucial where highest performance is the goal.
For special purpose application, this co-design is common practice.
For general purpose application, however,
a precondition for usability of a computer system
is an operating system which is both comprehensive and dynamic.
As no such operating system has ever been designed,
a sketch for a comprehensive dynamic operating system is
presented, based on
a straightforward hardware architecture to demonstrate how design
decisions regarding software and hardware do coexist and harmonise.

\vfill
\newpage

\tableofcontents
\newpage

\section{Origin}

To increase the performance of computers, a continual series of
design improvements has been implemented. Since physical limits
inhibit perpetual acceleration of single processing units, deployment
of multiple processing units in parallel was adopted.
There are various approaches to a generic solution, the ideal
being to solve as many distinct computational problems
as possible.

\subsection{Performance}

Computers have been invented to perform calculations automatically,
and faster than humans can do \lref{1936kz}.
Performance of computers has been increased continuously ever since.
\term{Moore's Law} postulates some linear gradient for continuously
increasing complexity of integrated circuits:
\cite{The complexity for minimum component cost has increased at a rate
of roughly a factor of two per year} \lref{1965gm},
later revised to \cite{doubling every two years} \lref{1975gm}.
However, complexity is not performance, and integrated circuits
are not computers.
Modern computers are made up of one or more integrated circuits,
with the basic paradigm in hardware design being unchanged
since the very beginning of digital electronic calculation machine history.
The architecture designed by J. Presper Eckert and John Mauchly \lref{2008gh}
-- widely known as \term{von Neumann architecture} \lref{1945jn} --
is still the base for most computers today (see \lref{1987tr}, p.417):
A single processing unit accesses a single main memory to store
both data and code.\footnote{The \term{Harvard architecture} (see \lref{2011hp}, appendix L) is
based on seperate memory for data and executable code.
It has often been used with DSP based systems to increase
memory access throughput. Compared to \term{von Neumann architecture},
performance at most differs in a factor of two,
a constant, so it is neglected subsequently.}

Still, in the early days, two substantially different categories
of computers have been developed. Precursors for large machines,
\term{mainframes}, are military projects for ballistic and cryptographic
calculations \lref{1973br}.
In contrast, microcontrollers have been developed to allow
increasingly complex algorithms in machine control units.
Later, mainframes have been miniaturised,
while microcontrollers have grown more complex,
and eventually both development paths have merged,
the Motorola 68000 -- introduced in 1979 -- being one of the first
processors powerful enough to run a work station,
yet suitable for embedded control projects.

Usually, the measure for performance
is specified in basic operations per time,
e.g. MIPS\footnote{The term MIPS is used in its original sense
-- \term{million instructions per second} --
throughout this paper.
However, MIPS as a measurement unit has often been misused,
e.g. by including heavy optimisation by the compiler into benchmark results,
thus pretending superior performance where in reality much
less instructions have been executed (see \lref{2011hp}, appendix L).
But even when MIPS is calculated according to its literal meaning,
counting the instructions executed per time, instructions differ widely
in their functionality from one processor design to another,
so two processors with equal MIPS rate may differ
substantially in the performance they deliver.
Even with all these considerations taken into account,
the MIPS rate is merely a theoretical upper bound,
as administrative tasks will consume part of the available
performance, and for generic applications it is rare to
exactly match the basic design of a computer system.}
or FLOPS (floating point operations per second),
where a basic operation is equivalent to the execution
of some processor machine instruction.
Given ideal conditions, these processor performance values
are possibly matched by overall system performance.
Ideal conditions may be found where computers are used for
highly specialised tasks, such as super computers built for
number crunching, or embedded controllers in automation.

With general purpose computers, though, things are different,
as usage conditions vary widely.
Certainly, application performance for several scopes has increased,
e.g. achievable quality of animated video sequences synthesised
in real time is much higher than some fifty years ago.\footnote{The
first computer animated motion picture ever has been
synthesised on the soviet {\cyr BE1SM}-4 in 1968:
{\it http://www.etudes.ru/ru/mov/kittie}}
On the other hand, for several areas, increase in application
performance is much lower than increase in processor performance
over the same period of time.
E.g., hypertext rendering carried out by a web browser takes about the
same amount of time now as it did some ten years ago, although
processor performance has increased substantially.
Even worse, there are tasks that take substantially
longer on some current personal computer than the equivalent
operation took on an average late seventies home computer:
While booting an Apple II computer took some half second,
whereof most of the time the machine spent generating a sound,
time to operational ready state (i.e. login prompt) for a
current work station may well be some half minute.

\subsection{Limits}

This is where \term{May's Law}, though not quantitatively proven,
turns out to be true:
\cite{Software efficiency halves every 18 months,
compensating Moore's Law} \lref{2007dm1}.
\footnote{Actually this observation is not new, it is known as
\term{Jevons' paradox}:
\cite{economy of fuel leads to a great increase of consumption}, and
\cite{an improvement of the {\normalfont [steam-]} engine,
when effected, will only accelerate anew the consumption of
coal} \lref{1866wj}}
One might argue this is polemic,
but even then software complexity
and performance requirements will increase,
not in implementation,
but for algorithmic and quantitative reasons.

As long as the basic architecture of computers is not changed,
continuously growing application performance needs can be
compensated by increasing processor and memory performance.

Unfortunately, there are physical limits:
\cite{Because information cannot travel faster than the speed of light,
the only ways of performing a computation more quickly are to
reduce the distance information has to travel, or to move more
bits of information at once. Attempts to reduce distance are
eventually limited by quantum mechanics} (see \lref{1991tw}, p.5).
Moving more bits at once in a single memory computer
increases the amount of die area needed for data bus connections,
thus further limiting the extent of functionality.
Not only is there a minimum size for structures to work,
but also a lower limit for production of these structures,
though new lithographic processes are able to push this limit further.
Thermal dissipation is a problem, where energy consumption is
concentrated in circuitry of descreased size.
These limits processor industry has approached
around 2003 (see \lref{2011pm}, p.3$f$).
Limits to processing speed are reflected by stagnation of
previously increasing core clock frequencies,
see \rfig{1}{t}{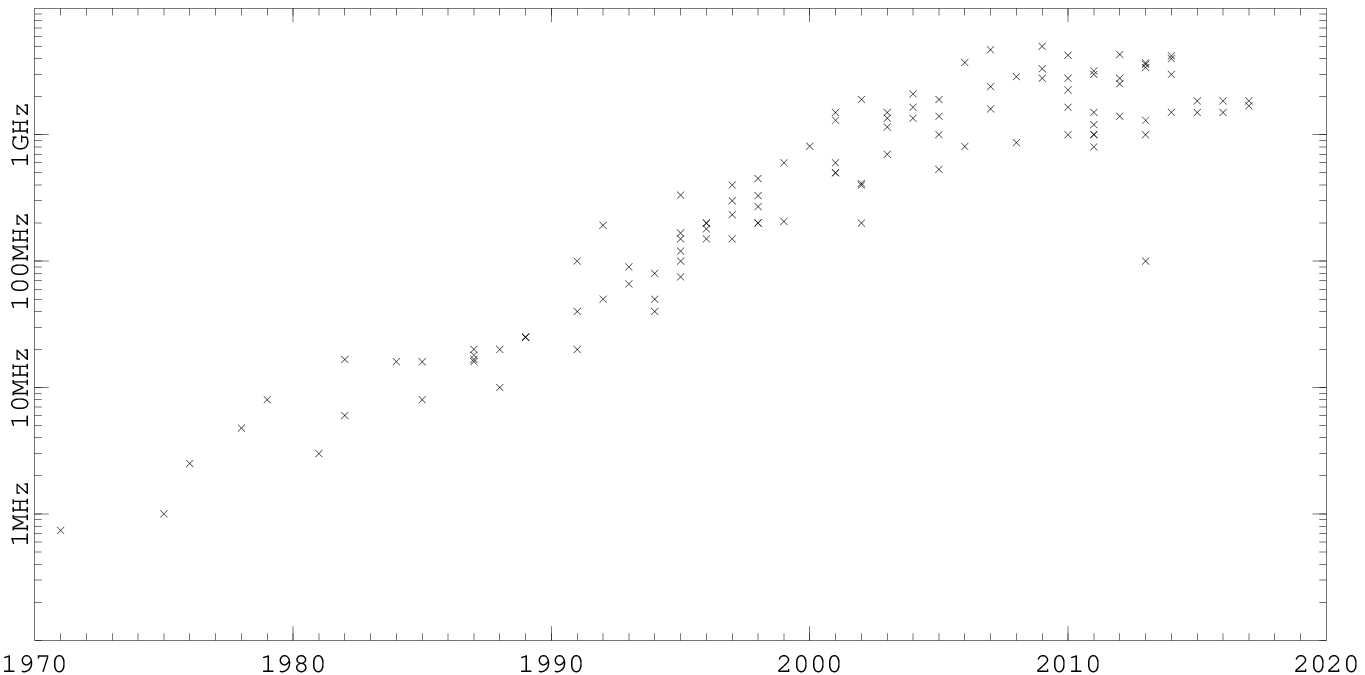}{: core clock input frequency
\footnote{The
figures are made up of the characteristic values of various
different architectures and processors, including
Intel 4004 and 8086,
Motorola 68020,
IBM POWER6+,
Freescale iMX6,
ARM Cortex-A7,
Intel XEON Phi 7290,
and Samsung Exynos 8895.}}.
However, an improved version of a processor with
a faster core clock frequency does not necessarily indicate increased
performance, as the same operation may need more clock cycles to
complete than with the previous design, e.g. along a pipeline.
Dividing the clock frequency by
the pipeline length reveals the true technical limit to completely perform
a single operation,
see \rfig{2}{t}{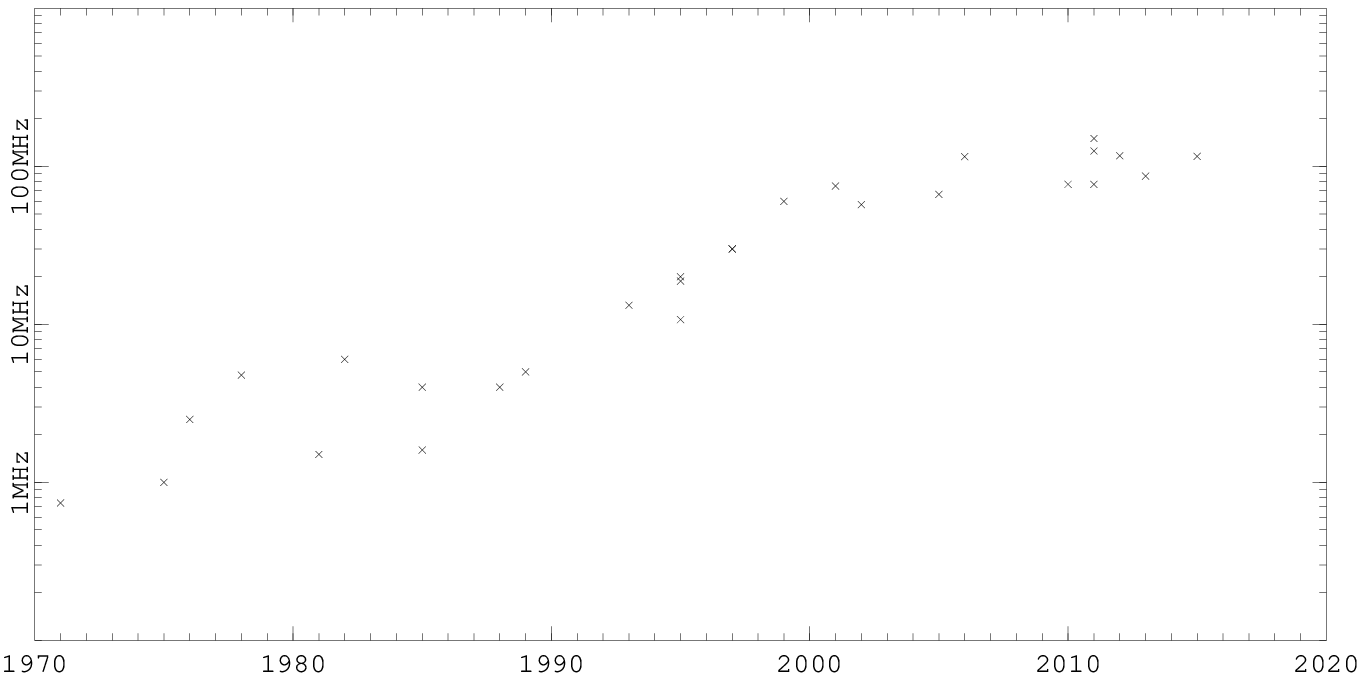}{: clock frequency per pipeline length $^5$}.
The thermal dissipation limits have been reached around
2005 (see \lref{1996hv}, p.55).
The limit in structure size ultraviolet light based production
has reached around 2012, using 32nm lithography technique,
see \rfig{3}{b}{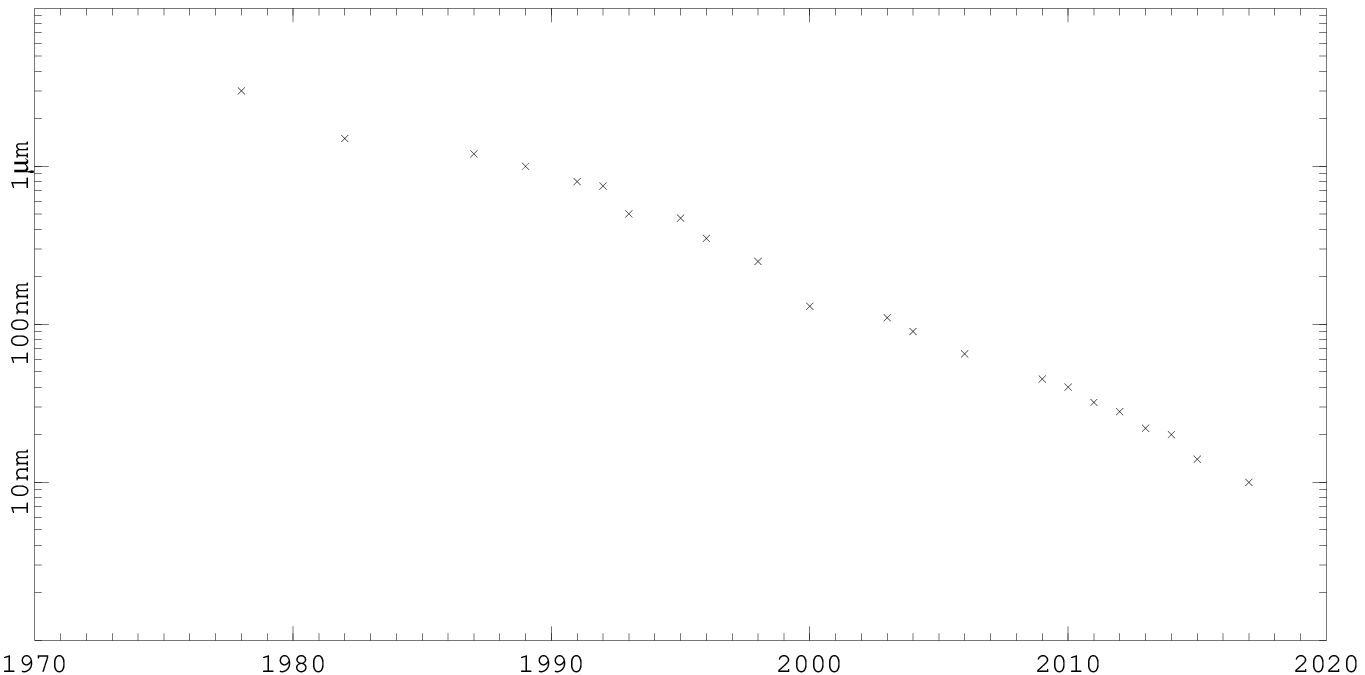}{: most advanced lithographic technology in use $^5$}.
Further reduction in structure size may be possible through advanced technologies,
e.g. extreme ultraviolet (EUV) lithography \lref{1998jb}\lref{2014bg}.
Eventually, these limits prevent further increasing core performance,
see \rfig{4}{t}{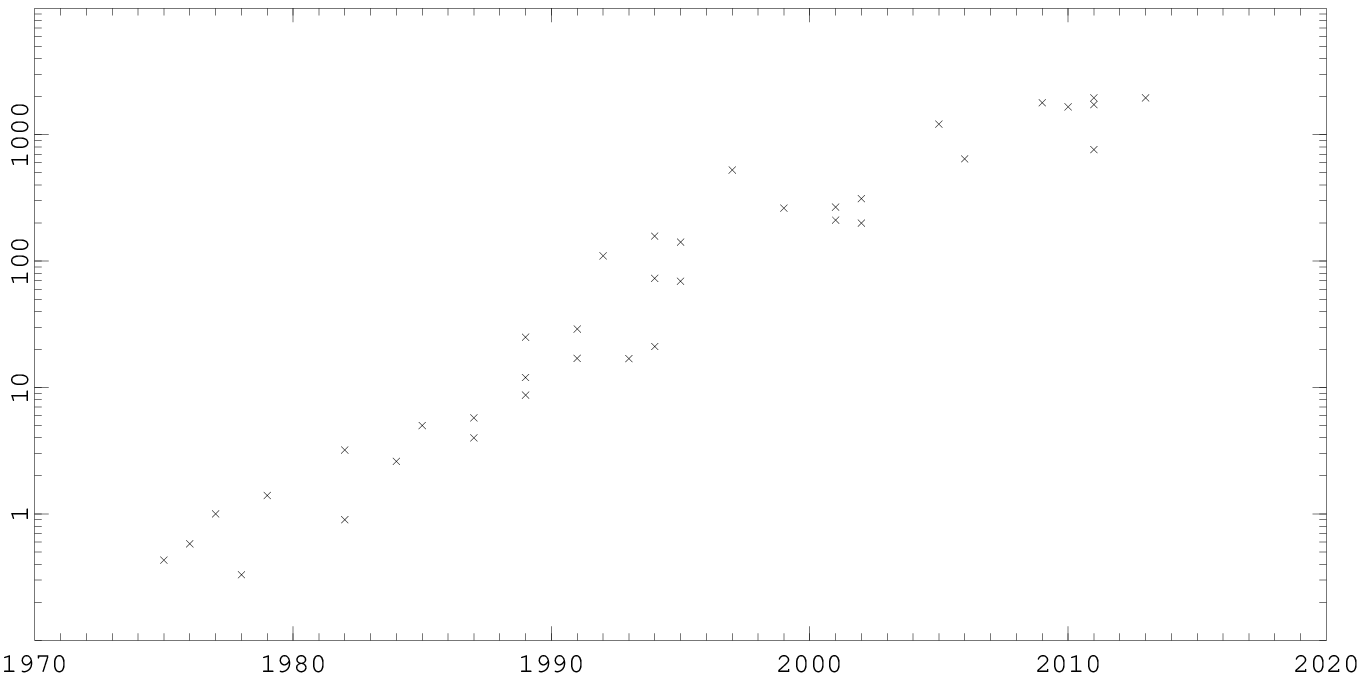}{: performance per core (MIPS) $^5$}.

\subsection{Transparent Structural Optimisation}

Additionally, up to this point,
various transparent structural optimisations have been applied,
none of which yields more than a fixed ratio improvement:

\begin{itemize}
\item Optimised instruction set encoding,
  shorten instruction words either generally (e.g. byte code)
  or for frequently used instructions (e.g. Xtensa, XMOS)
  to reduce memory access load.
  However, shorter instructions will encode less functionality
  and thus on average will result in longer instruction sequences.
\item The opposite approach to optimising the instruction set encoding
  is to use long instruction words (VLIW)
  encoding multiple operations per word,
  which at runtime are executed in parallel \lref{1983jf}.
  However, VLIW processors are not transparent from the point of view
  of the compiler, as it has to anticipate instruction scheduling
  to exploit the processors inherent parallelism.
\item Simplified specialised processing units (e.g. DSP, DMA),
  using less circuitry for a task,
  results in faster instruction execution.
\item Complex specialised processing units (e.g. FPU, GPU),
  accelerate distinct computations
  through higher degree of hard wired integration.
\item Instruction pipelining,
  splits up a single instruction execution into sequential stages,
  so different stages of successive instructions can be handled
  simultaneously.
  This way, it may happen that two subsequent instructions are
  executed at the same time, though not at the same stage of execution,
  with dependencies between these two instruction, e.g. the second
  instruction needs the result of the previous one as an input,
  before the result is readily available.
  A common solution for this \term{hazard} is to block execution
  of the second instruction (e.g. on Intel i860 \lref{1992in}),
  or more advanced, to add short cut
  circuitry for all possible constellations of instruction
  sequences to provide the following instruction with the needed input
  as fast as possible \lref{2007hh}.
  Another solution is to give up transparency,
  an early example being the Berkeley RISC processor \lref{1981ps}
  that defines delay slots for branch instructions causing the program
  counter to be updated only after the next instruction following the
  branch instruction.
  Transparency is also partially given up on Intel i860 \lref{1992in},
  where a floating point calculation is initiated by one
  instruction, but the result being ready at the end of the
  three stage floating point pipeline is stored into the destination
  register given by the third next instruction.
\end{itemize}

Specialised processing units and vectorised instructions
are special cases of an enhanced instruction sets.
More general approaches include the use of programmable
gate arrays to provide configurable hardwired,
and thus fast operations -- e.g. the
\term{instruction set extension fabric} (ISEF) on
the Stretch Inc S6000 processor \lref{2012hm}.

\subsection{Vector Processing}

To achieve higher data processing rates,
multiple data sets may be processed in parallel.
Either a single instruction sequence is performed
on all the data sets synchronously, which is called
the \term{single instruction multiple data} (SIMD) approach,
or each data set is handled independently from each other
by a distinct process, called the
\term{multiple instruction multiple data} (MIMD) approach,
according to \term{Flynn's taxonomy}\footnote{Flynn's
paper is about effectiveness of different computer designs at that time,
and what has been adopted as his taxonomy is only a fraction of
some classification described. As a sequel, other authors
state that Flynn's \cite{classification scheme ... is too broad},
it is \cite{a classification scheme by broad function rather than
a classification of the design} \lref{1988hj}.
Next to introducing \cite{Shore's taxonomy}
which subdivides only the SIMD architectures into different classes
-- and thus is equally incomplete --
they propose \cite{an algebraic-style structural notation, formalising
the functional units} \lref{1988hj}}
\lref{1972mf}.
What Flynn called MIMD is equivalent more or less to what today
is known as SMP (see section \ref{sub:symmetric}).
Additional architectures have emerged since, see table 5.

\begin{table}[htb]
\begin{center}
\begin{tabular}{|l|c|c|c|c|c|}
\hline
& & & MIMD & NUMA & \\
& SISD & SIMD & SMP & DSM & NoRMA \\ \hline
execution units & 1 & \multicolumn{4}{|c|}{$n$} \\ \hline
control units (cores) & \multicolumn{2}{|c|}{1} & \multicolumn{3}{|c|}{$n$} \\ \hline
memory units & \multicolumn{3}{|c|}{1} & \multicolumn{2}{|c|}{$n$} \\ \hline
address spaces (nodes) & \multicolumn{4}{|c|}{1} & $n$ \\ \hline
\end{tabular}
\end{center}
\begin{center}
{\it table 5: Computer architectures gross classification}
\end{center}
\end{table}

With \term{vector processing}, each data set is an array of data words,
the maximum length of the array defined by the hardware vector size.
Vectorised instructions perform the same operation on each single
word in a vector, where implementations differ in whether words are
processed simultaneously in an array of execution units, or
one after the other in a pipelined execution unit.
SIMD essentially is equivalent to vector processing (see \lref{1987wg}, p.322),
even for machines that are constructed differently,
e.g. the Connection Machine \lref{1985dh}.
\cite{SIMD are very good at some things, but inefficient at others} (see \lref{1991tw}, p.7),
because the design is
suitable only for algorithms that work on multiple data sets
synchronously in parallel using the same identical instruction
sequence for each of these sets.
Under optimum circumstances, this method does scale up to the
size of the vector, but for general purpose applications, this
is rarely achieved, because only a limited category of algorithms
lends itself to vectorisation.

\subsection{Asymmetric Multiprocessing}

Further, various asymmetric multiprocessing solutions
are used widely
-- though most of them usually are not referred to as such --
providing auxiliary processors for specialised tasks:

\begin{itemize}
\item direct memory access (DMA),
  an address driven data movement processor
  (e.g. Zilog Z8410 DMA controller \lref{2001zi})
\item floating point unit (FPU), and graphics processing unit (GPU),
  examples for complex special purpose coprocessors
\item digital signal processor (DSP),
  occasionally additionally available on-chip
  (e.g. Texas Instruments OMAP L138 \lref{2009ti})
\item other specialised coprocessors,
  e.g. the Cell BE featuring eight \cite{synergistic processing elements} \lref{2005cr}
\item peripheral control units,
  to handle various peripheral interfaces,
  occasionally even microcode programmable
  (e.g. PRU on Texas Instruments OMAP L138 \lref{2009ti})
\end{itemize}

\subsection{Symmetric Multicore}
\label{sub:symmetric}

About the same time industry reached
the limits in structure size reduction,
development of multicore processors started to fill the gap,
with the POWER4 by IBM being
\cite{the first non-embedded {\normalfont [multicore]} microprocessor}
commercially available around 2001 \lref{2011cs}:
Multiple processing units simultaneously execute
instruction streams one each.
In 2004,
Intel stopped single core development in favour of multicore \lref{2004lf}.

Designed for \term{symmetric multiprocessing} (SMP),
the computer is now capable of
independently executing multiple instruction sequences in parallel.
However, all the processing units
still access one single main \term{shared memory},
and thus constitute a \term{single node}.
\cite{Shared memory tends to become the governing system bottleneck
in cases where it happens that many processors try to access the
same memory at the same time} (see \lref{1987wg}, p.323).
There are strategies to reduce this problem,
large local \term{cache} memories being the most common,
and mandatory to reduce access to main memory itself (see \lref{2007hv}, p.22),
see \rfig{6}{b}{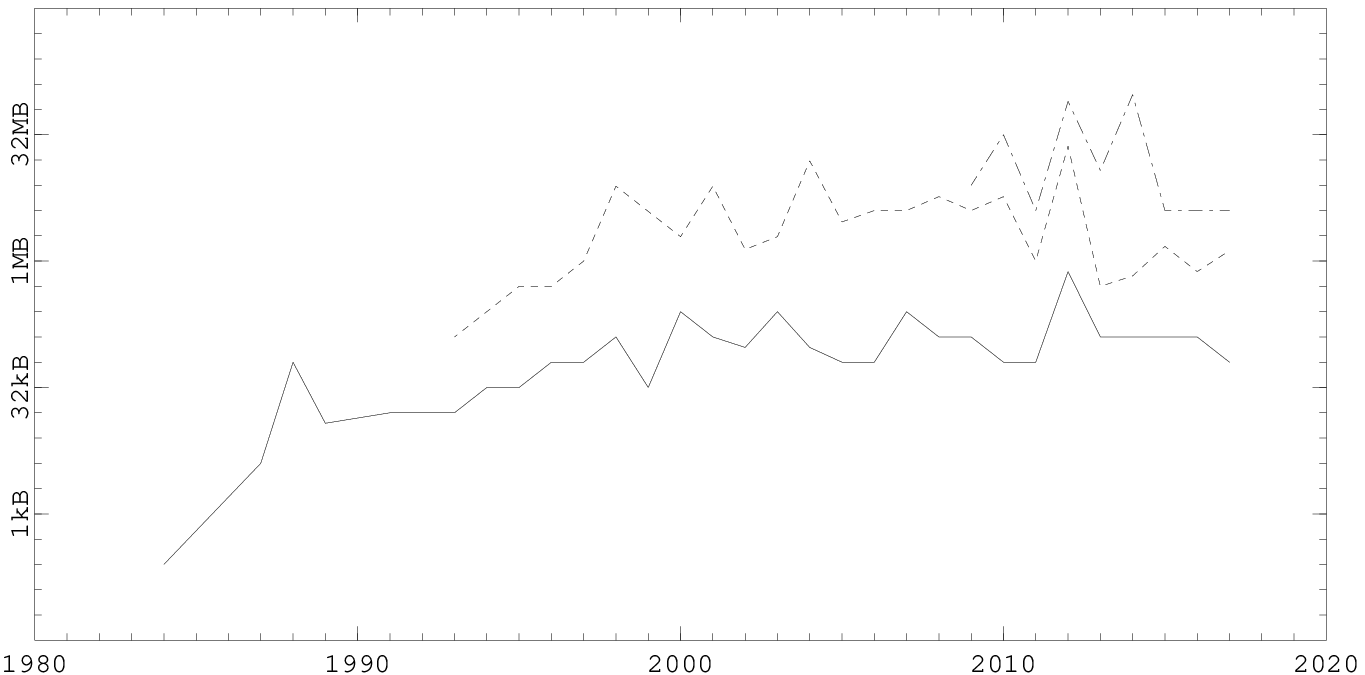}{: 32/64bit CPU caches size (1st, 2nd, and 3rd level) $^5$}.
But with the number of cores growing further
memory communication will increase further, too.
\footnote{Actually, cache memory has been introduced prior to
the advent of multicore systems: Clock rates for CPUs have
doubled about every two years, while access rates for DRAM
have doubled about every six years. Around 1985, rates have
been equal, but the discrepancy doubles every three years
since (see \lref{2007hv}, p.20).
With multicore systems, the problem is just worse.}

Besides, shared memory needs synchronisation methods to
resolve resource access conflicts,
e.g. locking through \term{semaphores}, or \term{transactional memory}. Both increase
system complexity, and do not scale \lref{2014yb}.

To optimise die area utilisation, cores may share some of their units,
e.g. the execution unit, to improve the benefit
from its stages at all times.\footnote{While
Intel calls this feature \term{hyperthreading},
other manufacturers call it \term{simultaneous multithreading}.
It is equivalent to the \term{skeleton processor} concept as
described by Flynn (see \lref{1972mf}, p.958)}
Apart from timing differences they cause,
shared execution units are transparent to software.

\subsection{Multinode Computer}

Single node computers are inherently ineffective:
\cite{Almost none of its billion or so transistors do any usefull
processing at any given instant} because most \cite{transistors
are in the memory section of the machine, and only a few of
those memory locations are accessed at any given time},
and \cite{the bigger we build machines, the worse it gets} (see \lref{1985dh}, p.4).

A measure for this imbalance is given by the
\term{capacity access time ratio}, which is the quotient from available
\term{storage capacity} in a system and the average \term{access time}
to it, a value that has increased -- and thus worsened --
by a factor of ten in less than
a decade ever since 1950,
see table 7.

\begin{table}[htb]
\begin{center}
\begin{tabular}{|c|c|c|c|c|}
\hline
year & 1950 & 1965 & 1980 & 2000 \\ \hline
bit/s & $10^7$ & $10^9$ & $10^{11}$ & $10^{13}$ \\ \hline
\end{tabular}
\end{center}
\begin{center}
{\it table 7: capacity access time ratio, according to \lref{2007hv}, p.22}
\end{center}
\end{table}

Whereas the multicore approach increases the ratio of
processing capacity against amount of memory,
it does not solve the memory bottleneck issue,
rendering the improved ratio largely useless.

Attempts are made to circumvent the single bottleneck
by splitting the shared memory into sections,
\term{distributed shared memory} (DSM),
providing
multiple, separate busses from the cores to memory \lref{2002cl}.
In theory, and for special applications, this approach
might mitigate the bottleneck, but not only introduces
it quadratic cost in interconnection\footnote{The
quadratic cost is worst case, and for the multinode
approach it is the same, but it is transparent, i.e. topology
handling to avoid access jam is not done by guesswork at
operating system level, but explicitely at application level
-- and thereby potentially much more efficient.},
it also hides the
memory topology from the software: Initially this might
seem an advantage, because it relieves software development
from being concerned with hardware details, but for
generic applications it is ineffective, as it is
not trivial for an operating system to optimally allocate
memory sections by predicting access patterns.
Research on how to further optimise allocation
largely concludes that user applications need to
support the operating system here \lref{2008bc}.
However, then allocation is transparent to the
application and there is no good reason to not let
the application completely assume control of data flow.

Provided no other transparent structural optimisations
or local parallelism approaches are invented,
to overcome the memory bottleneck,
one will eventually have to give up shared memory,
i.e. the single node multicore approach,
and handle communication among processing units explicitely.
With the \term{multinode} approach,
memory is split up into portions per core,
where each core has access to its local memory only,
and data exchange from one core to another is handled
explicitely using \term{communication channels} directly.

A number of systems have been designed according to
the multinode approach, namely Transputer based systems,
for an overview see \lref{1991tw}, p.8$f$, and p.354$f$.
Each of the multiple nodes usually follows the basic
\term{von Neumann architecture}
design concept (see \lref{1987bv}, p.4).
There has been discussion on which topology to choose
to connect the nodes \lref{1987wg},
until Inmos announced a packet switching based design
to overcome the restrictions of hardwired networks (see \lref{1991tw}, p.157).

Therefore, software design has to switch from
memory based data structure algorithms to
channel based data streaming algorithms.
Where application software is manually tailored to fit the hardware,
i.e. with special purpose computing, this is already largely the case,
and for many problems, suitable parallel algorithms are obvious,
such as searching and sorting \lref{1988gr}.

While being fundamental in super computing since 1972 \lref{1972bd},
apparently there is need for explicit parallel
implementations of average applications, too:
Originally designed to fulfil specific graphics computation tasks,
graphics processing units (GPU) today are used to implement
computationally intensive parts in scientific algorithms \lref{2004fm}\lref{2006dd}\lref{2015mv}.

For various problems, however,
such as system operation or code compilation,
parallel implementation may not seem straightforward initially.
To best facilitate software implementation,
\cite{a general-purpose concurrent computer must provide a simple way
for programs to be mapped on to the physical architecture of the
computer}, and \cite{this mapping must be achieved automatically if
portable software packages are to be written} (D.~May, in \lref{1989eh}, p.54).
The complexity of this automatic mapping will directly correlate
to the complexity of the entire system design, which thereby
causes the algorithms that are needed to fulfil this mapping to be equally complex.
This implies directly the need to keep the overall system design
as simple as possible.

However, it does not imply to hide the overall system design from
the application: For an application to benefit from a multinode
design it would be counterproductive to assume a different design
and let some underlying software layer translate access to resources as needed.
It definitely is no good idea to simply apply existing programming
models to a fairly different system design with least possible adaptions \lref{2006el}.
It is obvious that algorithms for a multinode design will look different
from those for single node computers, so porting tools to multinode
computers essentially means rewriting them.
Without doubt, this will be expensive, and maybe this explains
why \cite{we are still some way from having good standard toolsets
{\normalfont [...]} on parallel computers} (see \lref{1991tw}, p.355).
Without system software research responding to
fundamental progress in computer architecture,
though,
this progress will be of little value \lref{2000rp}.

\section{Review}

The predominating approach for concurrent computing today,
shared memory SMP, usually comprises a set of features
to optimise performance. These features need to be
reviewed for usefulness with the multinode approach.

\subsection{Shared Memory}

In a system with multiple nodes there is no need
to support multiple cores per node additionally on top of it,
because the scheme for inter-node communication
implicitely serves for inter-core communication as well.
Shared memory -- an extra scheme for inter-core communication --
would unnecessarily introduce additional complexity,
so it can be avoided altogether, resulting in a design
with only one core per node, the local memory model.
When a large number of nodes is provided,
then the amount of memory per node may be proportionally smaller,
as more nodes may contribute to the algorithmic needs
of some application.

\subsection{Cache}

Introduction of cache memory did help to reduce the negative
impact of the CPU-to-memory-gap, that emerged around 1985,
when CPU frequencies increasingly exceeded DRAM memory
access frequencies (see \lref{2007hv}, p.20$f$).
This gap already exceeds a ratio of 1:100 long since (see \lref{2008bc}, p.3).

As long as the combined memory, consisting of a slow memory
and a cache, is transparent to the core that accesses it,
using a cache does not cause harm.
This is almost always true for transparently addressing data,
but unfortunately it is not true for the timing behaviour of
the combined memory.
This may not be a problem for some applications,
but as soon as timing is an issue, the use of cache
memory spoils determinism.

In the latter case it might be preferable to deliberately
distinguish between direct access to fast local memory
on the one side, and explicitely accessing larger amounts
of slow memory on the other side.
This way DRAM is no longer the main implicitely addressed
memory resource, but fast local memory is.
DRAM, if needed at all, may be an external resource,
access to it handled by a dedicated process,
and consequently cache synchronisation,
bus snooping, and the like are not a topic anymore.

\subsection{Synchronisation}

As long as different processes are allowed to access
common resources simultaneously, some means of synchronisation
needs to be in place to avoid conflicts such as race conditions.

Without shared memory, however, the only common resources
available are distinct processes, access to which is available
through channel communication, which in turn provides implicit
synchronisation.
As a result, no further locking mechanisms need
to be implemented at all, neither in software nor in hardware.
This includes schemes like \term{semaphores} (see \lref{2001ed}) as well as
\term{transactional memory}.

For any external resource, one single process shall be responsible
for all access to it. This way synchronising access to it is
reduced to synchronisation in channel communication, which again
is implicit.

\subsection{Time-Shared Multitasking}

On systems with only one or a low number of cores, execution
of multiple processes is performed by means of time-shared scheduling,
using a single processing unit to run multiple processes,
usually alternating in a time-sliced manner.
Virtual addressing is used to both avoid memory fragmentation
and address conflicts,
and to pretend to supply more random access memory
than actually is available.
With message passing parallel computers,
the need for time-shared scheduling is not evident,
as long as there are enough cores
to provide each process with one core.
Otherwise, stale processes could be swapped out in
much the same way as memory portions of idle processes
on a single node multitasking machine are swapped out to disk.

With enough cores to provide one dedicated core
for each process in the system,
there is no need to share cores among processes.
Consequently, support for time-shared multitasking
is not needed.
It is abandoned altogether
in favour of dedicated core usage.

\subsection{Interrupts}

Usually, interrupts are used to handle external events asynchronously on
a core engaged otherwise.
This is necessary where the core is not fast enough to handle the events
through busy polling on a machine that does not provide a single
dedicated core for each single event,
or for a group of related events (see \lref{2001ed}, p.28).
The concept of interrupts had been introduced in 1957,
and it was immediately obvious it would wreck the processors
deterministic behaviour \lref{2001ed}.
On a machine with cores enough to provide one for each group of events,
and supposed these cores are fast enough, there is no need to
disrupt an executing process to handle external events.

To minimise latency from event occurence to its handling,
a wait instruction is introduced,
that allows the sequential execution of instructions
to stall until an event
-- out of a set of previously configured events for this occassion --
occurs (see \lref{2010dm}, p.25).

Note, that it is essential that the wait instruction is capable
of waiting for a set of events, not just a single event.
Since the core that handles the event is dedicated to exactly this task,
stalling the executing of instructions does not obstruct any other task.

This way, the concept of interrupts is no longer needed.
As a result,
since \cite{the interrupt mechanism turned the computer
into a nondeterministic machine with a nonreproducible behaviour}
(see \lref{2001ed}, p.13),
to abstain from it means to regain determinism.

\subsection{Exceptions}

Exceptions are used in an otherwise sequential
thread of instructions to handle -- as the name says --
exceptional situations.
This may be a special result of an instruction,
e.g. division by zero.
However, with instructions
always returning useful results no matter what the parameters were,
exceptions can be made superfluous. The division instruction for example
might take dividend and divisor, and return division result and remainder
and additionally some flag. Alternatively, the division instruction
might be implemented as some kind of branch
instruction, that branches on zero divisor.
Or it may simply return an undefined result on zero divisor,
assuming that case has been handled explicitely prior to division.

For channels that shall handle packetized data, i.e. besides
ordinary data words software needs to transmit and detect control tokens
-- e.g. \term{end-of-packet} token as used with SpaceWire \lref{2008es} --
reception of the right type of token may be anticipated
either by a special instruction, or by an additional configuration
option to the wait instruction: Wait for either type of token,
then jumping to different locations depending on the token type.

Non predictable or fatal exceptions (e.g. bus fault) might cause an
\term{exception message} be sent to an \term{exception handler},
which is another process (identified by its port number).
The former process may simply be stopped, asking for external
activity to handle the exceptional situation.

Eventually,
with no time-shared multitasking, no interrupts, and no exceptions
disrupting the control flow of a process,
context switches are superfluous altogether
-- except for the start of a new process
that needs an initial context setup.
To go without context switches not only simplifies
overall system design, but also saves significant runtime cost
where register files would have to be stored and loaded, and
cache tables flushed to avoid security flaws (see e.g. \lref{2018si}).

\subsection{Privileged Mode}

No context switches also means no context changes:
A process started in non-privileged mode can never
gain privileges: There are no syscalls.

The only way for a non-priviliged process to have a
privileged task done is to send a message to
a privileged process and ask it to perform the task.
Possibly, that latter process needs to check the
authorisation of the originator to decide.
The authority information plaited into the
originating port number could solve the problem:
The most simple case is a bit in the port number
indicating a privileged process sending from it.

\subsection{Peripheral I/O}

Peripheral I/O modules may be looked at as
auxiliary specialised processors -- either hardwired or programmable.
Where the core is fast enough to handle a peripheral interface
directly accessible through GPIO lines, there is no need for another
dedicated peripheral interface controller.
Latency is reduced using the wait instruction, to allow for
immediate response upon occurence of an event \lref{2009mm}.
There is no need for extensive peripheral circuitry,
except for just a few configurable shift registers.
Moreover, driver software is freed from the burden
to handle complex -- and often obscure -- subsystems,
as it now has direct access to the transmission lines.
This type of direct interface handling is good for
a large variety of interface types, including high speed
data transfer interfaces,
it has proven to cope with e.g. a 100~Mbit ethernet PHY on a 50~MIPS
core \lref{2012sb}.

\subsection{Asymmetric Multiprocessing}

Specialised processors increase the performance for special use cases,
usually executing in parallel to a conventional general purpose CPU.
This is asymmetric multiprocessing,
and for each specialised processor in a computer
special software needs to be written.
To decrease overall system complexity and cost,
specialised processors shall be abandoned in favour of
standard processors (see \lref{2007dm2}, p.10).

In a system with a single core per node and small local memory,
there is not much use for DMA, because most larger
data movements take place on inter-node channels.
For other examples -- DSP, FPU, GPU --
the main difference to a standard processor
are specialised instructions,
commonly floating point or other complex arithmetic.
It is not axiomatic to avoid instructions
that support special operations,
but by not restricting these to separate processing units,
the overall system design can be kept symmetric.

\subsection{Virtual Memory}

Shared memory accessed by multiple processes using
absolute addressing modes at instruction level requires
virtual address translation to avoid both
allocation fragmentation and address space conflicts.

However, these problems are not an issue on
a system consisting of large numbers of cores with
no shared memory,
each with its own small portion of local memory instead.
Memory allocation is process local:
A core and its local memory are an indivisible unit,
and from the system point of view memory is not
allocated, but cores are.

It may be desirable though to swap processes to optimise
locality in communication at runtime (see \lref{1985dh}, p.133$f$).
Hardware support for virtual channel addresses may prove
a useful feature in this context
-- i.e. determining and storing channel numbers at run time \lref{2013os}.
Furthermore, when resources are scarce -- i.e.
the system is out of cores, much like a shared memory system
is out of memory --
using a virtual channel address scheme makes it possible
to stall and swap out processes rather than memory pages,
eventually providing virtual resources in much the same way
as with a virtual address space that is larger than
the physical one.

Thus, simple flat memory process administration per node will be sufficient.

Still, multithreading may be implemented to a limited extent
to increase flexibility in resource usage,
which is similar to multiple cores sharing single stages
of the execution pipeline by passing access to the stages around,
thus avoiding stages to sit idle \lref{2009mm}.
To cope with memory fragmentation in this environment,
an instruction set restricted to relative addressing is used,
with all addresses relative to a set of base pointers,
e.g. instruction pointer, constant pool pointer, data pool pointer.
Avoiding absolute addressing altogether makes process
memory relocation a rather simple task.

\subsection{Power Management}

Power management refers to two different domains:
Internal power management, i.e. the core itself changing
to a state of reduced power consumption, and peripheral
power management, where peripheral interface controller
circuitry is partially or completely switched off when
not in use.

The latter is a topic only for external additional hardware,
as internally there are no peripheral interface controller
blocks. Switching off external circuits may well be left
to explicit handling through driver software.

Internal power management consists solely in reduced
clock frequency modes and thus reduced power consumption for a core,
that is stuck in a wait instruction \lref{2009mm}.
As this state is entered automatically, there is no
need for the software to take further action:
Power management is implicit.

\subsection{Cost}

Though not strictly necessary, it might be desirable to
have available some of the above mentioned features,
as they might simplify handling the machine.
However, by omitting unnecessary features the die area
needed for the implementation of a single core is reduced,
thereby increasing the potential number of cores per chip.

Another motivation to simplify the hardware design is
software engineering cost: specialised circuitry needs
dedicated software drivers, and additional hardware features
need to be handled by software, even where they are intended
to simplify overall system design.
Even worse, different features may coincide, e.g. power management
must be implemented again with every single peripheral driver,
multiplying the extra cost for both features.
As a consequence, potentially,
\cite{as the number of capabilities added to a program increases,
the complexity of the program increases exponentially} \lref{1970cm}.
Replacing complex dedicated peripheral circuitry by
direct access to the transmission lines may reduce cost
in driver development substantially.\footnote{E.g,
the XMOS ethernet driver of 400 lines of code compared to
a Linux kernel Intel e100 driver of 3200 lines of code -- both written in C.}

Simplicity in software design is not only good to save cost,
but also \cite{for reliability simplicity is an absolute prerequisite} \lref{1975ed}.
Likewise, for security reliability is an absolute prerequisite.
It is known that the number of bugs in large software systems
directly relates to the size of the software (see \lref{1979yc}, p.370).
On a shared memory computer, all system software on the single node
adds up, yielding much higher complexity than programs on isolated
nodes would do, hence
\cite{programming of a system in which the programmer has
explicit view of memory is much more complicated and error-prone
than the programming of message based systems} (see \lref{1987wg}, p.323).
Consequently, to support software reliability, it is inevitable
to keep the hardware design as simple as possible.

Moreover, choosing the most simple basic hardware design improves
software portability, simply by reducing the number of features
and special cases that would need to be handled porting the software.
As
\cite{the successful exploitation of concurrent computers now depends
more upon achieving software portability than upon any other
single factor} (D.~May, in \lref{1989eh}, p.54),
simplicity is crucial for the success of any basic system design \lref{1970cm},
just like \cite{a key property of the von Neumann architecture
for sequential computers is efficient universality} \lref{1989lv}.

\section{Feasibility}

How should a simple multinode design look like?
What are its dimensions,
and how does its order of magnitude relate to feasibility?

\subsection{Sketch}

From the collection of considerations it can be concluded,
that three basic conditions hold for the design of an improved
hardware architecture:

For one, to overcome the \term{active silicon imbalance} \lref{1985dh},
it must provide as many processing units as possible.
Because a SIMD architecture does not match arbitrary algorithms,
and because the shared memory approach does not scale,
it must be a message-passing multinode architecture.

Further, the design of the individual processing units, the cores, must be
as simple as possible. In reducing their size in hardware
implementation the highest possible number of cores
will be available per chip. Furthermore, reduced complexity
decreases cost for software development, maintenance,
and portability \lref{1996ed}.

Lastly, the software used with it will
\cite{dynamically allocate and deallocate processors in the same way
that a sequential program dynamically allocates and deallocates
memory} \lref{1989dm}. %p.32
Because
\cite{dynamic allocation requires coarse grain parallelism} \lref{1989dm}, %p.32
instruction wise forking and synchronising execution is not suitable.
Instead, parallelising algorithms into separate processes
matches the concept of \term{allocatable cores},
demanding channel based message passing.
This essentially is \term{communicating sequential processes} \lref{1978ch}.

Because software will allocate processing units instead of just memory,
thus starting processes, a program consists of an arbitrary,
possibly variable or even unbound number of processes.
This is analogous to a single node computer program allocating
memory pages at will.
As a consequence, the individual processes cannot allocate
more memory at runtime,
but have their static amount of memory assigned at process start time,
together with the core to run on.
Processes that need to allocate memory dynamically will have
to allocate cores instead, running supportive processes,
which in turn may or may not provide more functionality than just additional memory.
This way, memory is not some passive resource, it is active memory.
For implementation reasons, the amount of local memory the
process has available, will be quite limited -- to avoid
undermining memory locality -- but not necessarily fixed to
some system wide constant value.

To achieve variable local memory size a region of common
local memory may be shared among a set of processing units.
This solution is implemented in XMOS XS1 \lref{2007dm2},
but it reintroduces
all the disadvantages of shared memory.
To reduce impact on software reliability, an MPU may be added
to prevent access of one process to the memory section reserved
to another process.
E.g., the \term{Null Operand Parallel} processor implements implicit
range checks to trap memory access violations \lref{2016os1}.
An advanced solution would be an MMU to assign the required number
of memory blocks to a process at startup time. In conjunction with
an externally driven (e.g. through the interconnect)
process startup circuitry,
it renders local software to control memory assignment superfluous
(see section \ref{hwsupportsched}).

While instruction pipelining applied to a single core introduces
either wastage or hazards,
this is not the case where the execution pipeline is shared
among several cores, each
of the cores using only one stage of the pipeline at a given time.
To sustain reliable process response time
-- and thus not loose the ability to handle port based peripherals --
timing with the shared execution pipeline must be predictable,
e.g. by using a deterministic round robin scheduler.
The XMOS XS1 implements this restricted variety of a hazard free
shared execution pipeline.
Combined with fully hardware controlled memory assignment,
the aspect of sharing memory and computational resources
among multiple processes is no longer visible to software at all,
eliminating any need for software supported resource conflict
management.

Besides \cite{many processors} the second basic requirement
is \cite{programmable connections} (see \lref{1985dh}, p.14$f$),
as without programmable connections
\cite{the algorithm is designed to suit a particular configuration.
  This is satisfactory for embedded applications, where the
  configuration can be determined by the application.
  It is obviously unsatisfactory for a general-purpose computer} (D.~May, in \lref{1989eh}, p.54).
This requirement is easily met by programmable interconnects
as found e.g. on XMOS XS1 processor chips \lref{2009mm}.

\subsection{Size}

The size of the execution unit that will be used in a parallel system
design is as yet unknown, because no dedicated solution has been
implemented in hardware so far. To obtain an estimation,
numbers available for comparable designs are investigated, see table 8.
Further it is assumed that for the execution pipeline
a simple design of four stages is chosen, and that only a fraction
of all resources will ever be active. Following this consideration,
it is acceptable to plan one execution unit to be shared
amoung eight processing units each.

\begin{table}[htb]
\begin{tabular}{|c|c|c|c|}
\hline
design & transistors & logic cells & local memory bits \\ \hline
J1a \lref{2015jb} & & 1200 & 64k \\ \hline
Motorola 68000 & 68k & & - \\ \hline
Inmos T800 & 300k & & 32k \\ \hline
(OpenSPARC T1) \lref{2013mf} & & 285k & 192k \\ \hline
\end{tabular}
\begin{center}
{\it table 8: compact microprocessors key figures}
\end{center}
\end{table}

Deduced from these numbers,
it is assumed that a suitable processor, including
execution pipeline, register banks, interconnect,
and control logic, designed
for shared use with eight processing units, will
require no more than one million transistors.
It is further assumed that a major portion of these
transistors represents storage facilities,
like register banks or routing tables,
so half of it may be replaced by block RAM
in an FPGA based implementation.
To further simplify the estimation, 15 transistors are taken as
an equivalent of one logic cell (see \lref{1992bf}, p.22).
Assuming all local memory is CMOS SRAM,
requiring 6 transistors per bit, plus fringe electronics,
subsequent estimation is based on 8 transistors per bit in total.
It is undisputed, that these assumption are very vague
and suitable only for a rough feasibility estimation,
particularly, because the number of transistors alone does
not linearly correspond to the die area the design would
need, as is explained comprehensively in \lref{1989dm}.

Recent advances in microprocessor manufacturing have
shown it possible to cram 4 billion transistors onto
a single processor chip \lref{2016bb},
see \rfig{9}{t}{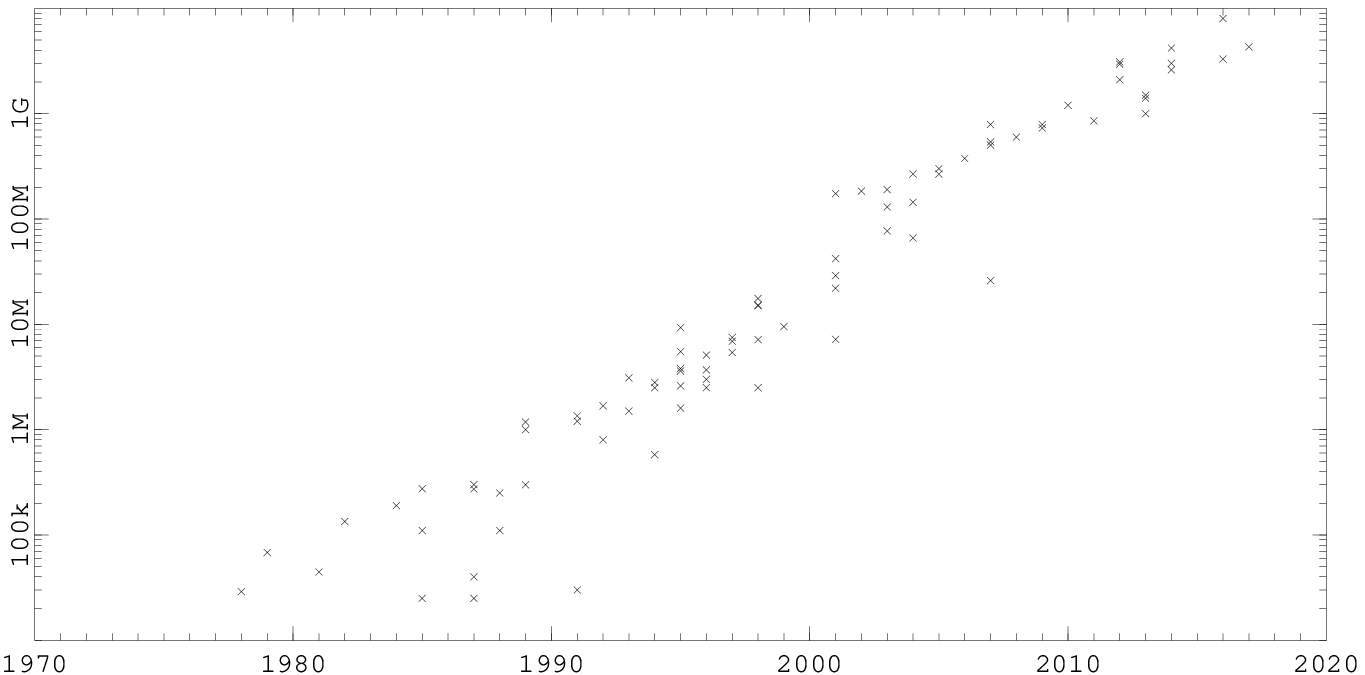}{: chip circuitry size, CPU transistor count $^5$}.
While this may serve for an estimation
of what can be achieved in theory, for an estimation
concerning some early hardware prototype the characteristics
of the largest currently available FPGA are taken as a basis:
The Xilinx Virtex-7 XC7V2000T offers some 1954k logic cells
and 46512kbit of block RAM \lref{2017xi}.
Based on these numbers,
\rfig{10}{htb}{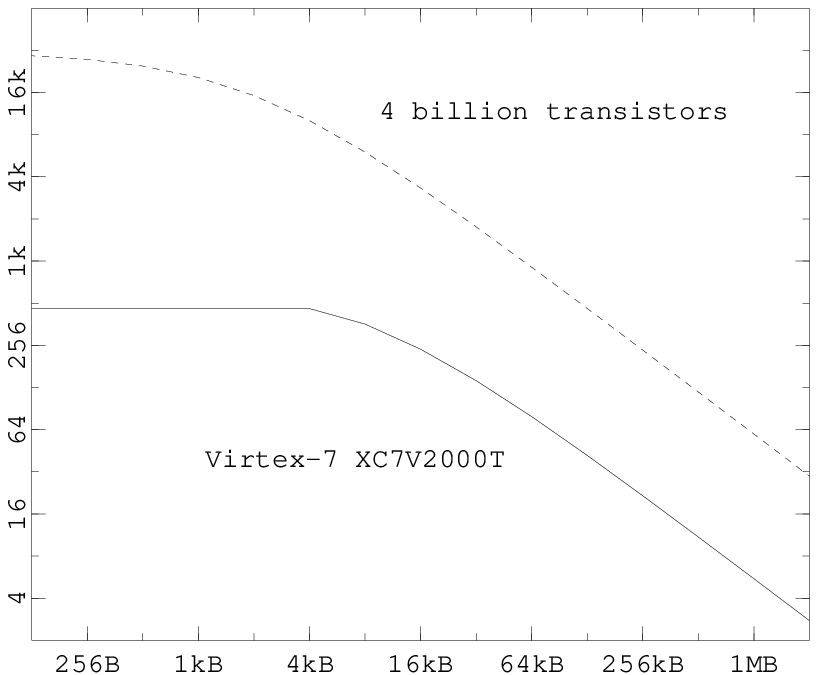}{: cores per chip, depending on local memory per core}
shows a rough estimate for the number of cores that might fit on one chip,
as a function of the size of local memory per core.

Further limits, like heat dissipation or needed number of pinout,
have not been considered so far.

\section{Operation}

Cultivation of multinode computers for general purpose use,
i.e. to run a wide variety of applications on it, asks for
abstraction to a degree so as to relieve application software
programming of considering hardware details in too much depth.
The overall multinode design should not be hidden from
the application, but quantitative characteristics should -- like
number of cores available, or the underlying network topology.

\subsection{Abstraction and Management}

An operating system is defined to be some basic software
executing on a computer,
providing arbitrary application software with resource management,
and ideally with full hardware abstraction \lref{2009sg}.
This operating system turns the specific hardware
implementation into a general purpose computer system,
and allows to design application software
in a hardware independent way.

Hardware abstraction is meant to standardise access
to peripheral devices.
Hardware abstraction in the sense of portability to a different
type of computer is not the task of the operating system
but largely the task of compiler tools.
When coping with parallel computers,
the question of how to organise parallelism is independent
of peripheral devices, so the latter may be left out of
the operating systems kernel part.
Nevertheless, peripheral devices need to be controlled,
so a set of processes for this purpose may be considered
part of the operating system.

Resource management is to fulfil an applications
demand for processing capacity and memory space,
and this is where parallel properties of a system
inherently determine the algorithms to use.

As already described in \lref{2013os},
on a single node computer,
system global memory is the main resource to administer,
and it usually is portioned in memory pages,
see \rfig{11}{t}{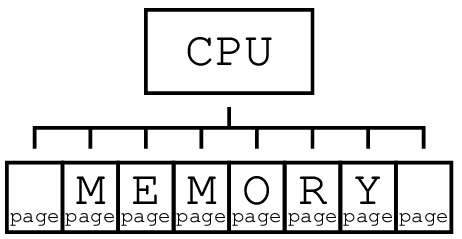}{: global memory is a page-wise resource}.

On a multinode system with a single core per node however,
the \term{processing units} are the main resource,
i.e. the cores, which constitute computing capacity together with local memory,
see \rfig{12}{t}{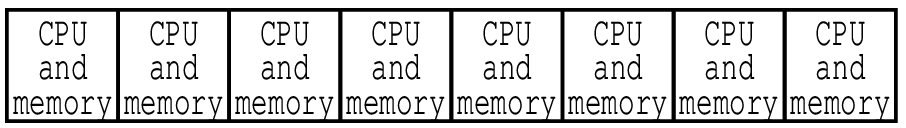}{: local memory is part of a core-wise resource}.
Allocating dynamically these units instead of bare memory has been
suggested earlier:
\cite{processing resources ... be allocated and deallocated as freely as memory}
\lref{1989dm}.

The part of the operating system controlling processing
capacity usually is called the \term{scheduler}.
There are two purposes it may serve:
One is \term{arbitrary process launch}, i.e. starting
the execution of a random new process at system runtime.
The other purpose is \term{multitasking},
i.e. controlling the execution of multiple processes
at the same time. Not all operating systems do serve
both purposes. On large computer systems in the 1960-80s
common practice was \term{batch control}: There was no
\term{multitasking}, but full processing capacity was allocated
to a single process, and the next one would be started only when
the previous had finished executing \lref{1974as}.
In embedded control, \term{static schedulers} are quite common,
that do not support \term{arbitrary process launch}.

On computers that do not provide a dedicated core
for each process to execute, multitasking is achieved
by time sharing the processing capacity, through e.g.
multiprogramming (see \lref{2001ed}, p.20).

An operating system is to be called \term{dynamic}, when
it supports starting random numbers of new processes
to work in parallel (or in a pseudo parallel manner)
at runtime, i.e. when it supports both \term{multitasking}
and \term{arbitrary process launch}.

An operating system is to be called \term{comprehensive}, when
its resource management provides unified access
to the full range of processing resources.
Only then the computer can be utilised by standard applications,
and thus independent of some front-end processor.
E.g., for GPU accelerated systems this is not the case,
as there is no native operating system support,
but some library interface at application design level
for explicit resource usage.

\subsection{History}

Computers being commercially available at first in the 1950s
were expensive highend special purpose calculation
machines, their market section being comparable to super
computers today, but they were not supplied
with an operating system.
Based on the then new concept of interrupts, D{\ij}kstra introduced
concurrent control of peripherals to the Electrologica X1 in 1957, but there
was neither multitasking nor dynamic job allocation (see \lref{2001ed}, p.15$f$).
Though modern super computers may have some kind of
operating system, usually it is not comprehensive,
and definitely not suitable for average work stations.

Meanwhile, electronic components have been used in automation,
simple processing units and microcontrollers complementing the
range of parts used subsequently.
Only hardware abstraction has been provided in this area for long,
and static scheduling is still a common approach today.

It took until 1969, when Unix was developed,
and subsequently became the first widespread
comprehensive dynamic operating system for
single node computers.
Some thirty years later, in the late 1990s,
derivates of Unix started to be used in automation,
enabling unified software development
for the two areas of computation machines,
work stations and embedded control \lref{2013cs}.

Parallel computers based on message passing
have been developed since the 1970s \lref{1991tw},
none of these being equipped with
a comprehensive dynamic operating system up to now.
In automation, systems consisting of thousands
of microcontrollers passing messages
to coordinate activities (\term{signalling}) are quite common,
digital telecommunications being a prominent application
(see e.g. \lref{1990lt}).
However, these systems are not equipped with
comprehensive dynamic operating systems, either.
Yet, regarding static scheduling, it has been described earlier as
\cite{natural to remove these restrictions}
in favour of \cite{dynamic resource allocation} \lref{1989dm}. %p.27

Up to now,
explicitly parallel algorithms account for only a few
specific parts of software, whereas
\cite{the two main arguments against the use of parallel computers
  today are that they lack software, and that they are difficult
  to program} (see \lref{1991tw}, p.9).
Consequently, and because
\cite{the future of parallel computation may be strongly influenced
  by the extent to which efficient universality can be found and
  harnessed} \lref{1989lv},
implementing the basics for some parallel operating system
is the precondition for any application being implemented
explicitly parallel in its entirety.

\subsection{Related Work}

There have been efforts to develop such an operating system.

\term{Amoeba}
was designed as a Unix based multi machine cluster,
i.e. running a full multitasking operating system on each single node
\lref{1990tr}.

\term{Barrelfish}
is not based on an existing operating system,
but is designed for use with conventional hardware (x86\_64 or ARM based),
that provides large amounts of random access shared memory per processing unit,
430kB for the CPU driver alone \lref{2009bb}.

\term{Plan 9} is a distributed operating system,
designed for use with conventional work stations,
and ported to a variety of processor architectures \lref{1991pp}.
It is a consistent further development of the basic concepts of Unix.

\term{Helios} was an operating system designed specifically
to run on multinode hardware, namely Inmos Transputer (see \lref{1991tw}, p.297).
Parts of its design are similar to the one presented here:
The Process Manager and the Loader cover tasks of
the scheduler, the dispatcher, and the loader
(see section \ref{sec:implementation}), but provide
a variety of further functionality, like signal handling
and real time clock control (see \lref{1991ps}, p.20).
However, the minimum amount of memory per core is
one megabyte (see \lref{1991ps}, p.14),
limiting the total number of cores in a
reasonable system setup to some thousand.
Furthermore, parts of the operating system refer
directly to the design of some frontend computer,
so \term{Helios} is not comprehensive.
Nonetheless, the overall design of \term{Helios} has been
quite promising at its time.

The name \cite{Helios} matches the idea of a distributed
operating system quite well, so it is not too surprising
that there are more projects that bear the same name:
\term{Helios} was a research project introducing
satellite kernels in heterogenous multiprocessing \lref{2009nh}.
Its memory consumption exceeds 32MB per node (see \lref{2009nh}, p.3).

\term{Vortex} is an experimental \cite{event-driven
multiprocessor operating system} run on Intel SMP hardware \lref{2003kj}.

\term{Corey} is a many core operating system
research project experimentally run on 16 core
x86\_64 based SMP hardware with megabytes of cache available \lref{2008bc}.
Numbers on memory footprint are not available, except that kernel
source exceeds 11000 lines of code, most of it written in C.

An exemplary list of five purposes of an operating system
for multinode machines, as it is still widely accepted, is given
in \lref{1987bk}, p.208:

\begin{description}
\item \cite{Multitasking} -- is obsolete where each user process
is assigned its dedicated core.
\item \cite{Channel multiplexing} -- shall be done in hardware.
\item \cite{Memory management} -- will not be needed as access for each process
is restricted to the local memory of its core. The idea that a
\cite{process must have access to more memory space than physically
attached to one processor} is a conclusion from the wrong
assumption that a complete application program needs
to run on one single node, or utilising shared memory
employing global address space.
\item \cite{Load balancing} -- is a task indeed, though a secondary one,
when it comes to optimising a given design.
\item \cite{Support of the basic data structures: ...
Garbage collection for heavily parallel machines is not solved yet.
Therefore it seems simpler not to share expressions among different processors} --
this is very true, and as allocation refers to processing units,
rather than memory, these processsing units would be the items to collect.
\end{description}

Suggesting the invention of a \cite{new model of computation}, Sterling
proposes a \cite{new co-design cycle of all levels of the system
software and hardware}, explicitely including the \term{operating system}
\lref{2009ts}.
This co-design is surely important, but there is not much
sense in asking for the next model of computation, when for
the second last model -- \term{communicating sequential
processes} -- that co-design has not yet been
performed. However, the paradigms to
achieve this new model have each already been introduced separately,
so all that is needed is to combine and implement them
consequently, particularly a comprehensive dynamic operating system.

\subsection{Requirements}

To be utilised with a comprehensive dynamic operating system,
a parallel computer system needs to fulfil some minimum
requirements in size:

\begin{description}
\item -- The size of local memory per core must suffice to hold
executable code of a single process and the amount of data it
needs to store locally.
\item -- The number of cores must suffice to run all programs
-- system functionality and user applications --
in parallel, where each program consists of a number of processes,
and thus needs the corresponding number of cores.
A lower limit for the number of processes that make up a program is
given by the overall memory need divided by the core local memory size.
\end{description}

The size of the executable code of a single process
is comparable with the size of a subroutine in sequential programming.
On average, it will be some kilobyte.

On current single node computers,
four kilobyte is a convenient basic memory allocation size,
called a \term{memory page}.
It is directly comparable to the basic allocation unit
of a parallel computer, a core and its local memory.

Both estimations combined imply that some eight kilobyte will
be a suitable average size for a single cores local memory.
Note that this differs by orders of magnitude from the gigabytes
per core todays work stations provide.

Average single node work stations run some two hundred programs
in parallel -- though most of the programs are inactive most
of the time.
Programs substantially differ in memory size,
simple system tools will do with just a few kilobytes,
while large office applications may well use megabytes.

Instead of counting memory sizes of single sample programs,
the overall memory utilisation
of a single node work station is found to be some gigabyte.
Division by eight kilobyte -- the local memory size per single core --
results in some 128 thousand cores per computer.

These numbers match research on parallel computing in the late 1980s,
which has shown it useful to think in the range
of \cite{at least tens of thousands} (see \lref{1991tw}, p.5)
up to \cite{a million processors} (see \lref{1985dh}, p.5, and \lref{1989dm}, p.36),
recent publications affirming that it is realistic to project
\cite{millions of processors} (see \lref{2007dm2}, p.2).

\subsection{Consequences}

As explained above,
designing a massively parallel computer for
general purpose use directly implies the need for an
appropriate operating system to be implemented.
On the other hand, implementing an operating system
to be used with massively parallel computers
requires hardware to run it on, be it real or simulated.
Consequently, design of both hardware and software
should be carried out simultaneously,
comprising the following main tasks:
\begin{description}
\item -- Design and prototyping of adequate processor hardware
\item -- Engineering of a parallel computer hardware
\item -- Design and implementation of an operating system
\item -- Implementation of application software
\end{description}

Realistic project scheduling
asks to select the bare minimum from this list,
i.e. use existing or simulated hardware,
focus on the basics of an operating system,
and the minimum set of tools to control it.

\section{Implementation}
\label{sec:implementation}

\subsection{Hardware}

An incomplete list of multicore processors and computers
gives an impression of the differences in local memory size,
see table 13.

\begin{table}[htb]
\begin{small}
\begin{tabular}{|p{1.3in}|p{.8in}p{.7in}|p{.9in}|p{.7in}|} \hline
architecture & $\mu$P & & local memory \newline per core & cores per \newline computer \\ \hline
Greenarrays GA144 & F18A & \lref{2011ga} & 64 word & 144 \\ \hline
Parallax & P8X32A & \lref{2011jm} & 2 kB & 8 \\ \hline
XMOS XC-2 & XS1-G4 & \lref{2009mm} & 8 kB & 32 \\ \hline
Tilera & TILEPro64 & \lref{2013tc} & 64 kB\footnote{on-chip L2 cache per core} & 64 \\ \hline
IBM PowerPC & POWER8 & \lref{2016bb} & 512 kB$^{11}$ & 96 \\ \hline
Ambric & AM2045B & \lref{2006th} & 1 kB & 344 \\ \hline % asymmetric SR vs SRD, discontinued 2012
Kalray & MPPA2-256 & \lref{2016ki} & 8 kB & 256 \\ \hline % plus 24 cores asymmetric
Adapteva & E64G401 & \lref{2012ai} & 32 kB & 64 \\ \hline
\end{tabular}
\end{small}
\begin{center}
{\it table 13: multicore processors key figures}
\end{center}
\end{table}

Both F18A and P8X32A do not provide enough local memory per core.
The number of cores on P8X32A is too low for even the minimum
operating system test setup.
TILEPro64 does not provide hardware based channel communication means.
AM2045B has been discontinued 2012.
For MPPA2-256, detailed information or an evaluation board is not publicly available.
With 32 cores, XS1-G4 is suitable only for basic operating
system tests, but it provides hardware based channel communication,
and an appropriate amount of local memory per core.
It lacks details in channel synchronisation
-- checking channel data availability is only implemented on channel input,
so non-blocking output would need some software work-around --
but basic tests can be done without.

The XMOS XC-2 computer was chosen for a first prototype,
but this first approach has been cancelled later,
because of the deficiencies described above,
mainly inavailability of test hardware with more than 32 cores.

Operating system basics have been implemented, the concept
being similar to the one realised with the second prototype
(see section \ref{sec:simulation}), the main differences are:

\begin{description}
\item -- No boot code is implemented, toolchain provided initialisation is used instead
\item -- Software is written in plain C \lref{1978kr}
\item -- Ethernet driver is implemented with IP/UDP \lref{1980jp}
         and TFTP\break protocol \lref{1992ks}
\item -- UART driver and console
\item -- No application software except
         a simple command interpreter and
         a simple system state inspection tool
\end{description}

Arbitrary process creation has been proven to be possible,
and the concept of generic hardware based peripheral access
has been verified and shown it possible to achieve
full 100 MBit throughput
on ethernet \lref{2012sb}.

Both implementation and tests also showed, that the overall
concept of channel based communication between processes
is suitable.

\subsection{Simulation}
\label{sec:simulation}

To reduce the dependency on available hardware,
a second prototype has been based on simulated hardware.
For this purpose, a very simple processor has been designed,
the \term{Null Operand Parallel} processor \lref{2016os1}.
Parts of its design are inherited from the XS1-G4 concept,
namely hardware channel communication support and
fixed round robin thread scheduling.
For properties not relevant for the prototype the
most simple solutions were chosen to reduce overall
project complexity, e.g. by using a byte code based design.

To allow the simulated processor to be programmed using
a high level language, a compiler is needed,
and because it needs to be implemented anew anyway,
a customised programming language
-- the \term{Guarded States Language} --
has been presented,
which combines concepts from different existing languages
to support programming the prototype most smoothly \lref{2016os2}.

No peripheral drivers have been implemented,
except for a simulated text console
and an interface to allow reading and writing files
from outside the simulated environment.

The processor implements four nodes
-- each capable of executing eight instruction sequences in parallel,
i.e. eight cores per node -- and one common switch
for channel communication both internally between the
four nodes and externally to other processors.

A simple test setup involves e.g. 16 simulator instances,
each of which simulates a processor consisting of four nodes and a switch.
This sums up for a total of 512 cores.
The single simulators are connected via simulated external links,
i.e. data sockets
(\rfig{14}{t}{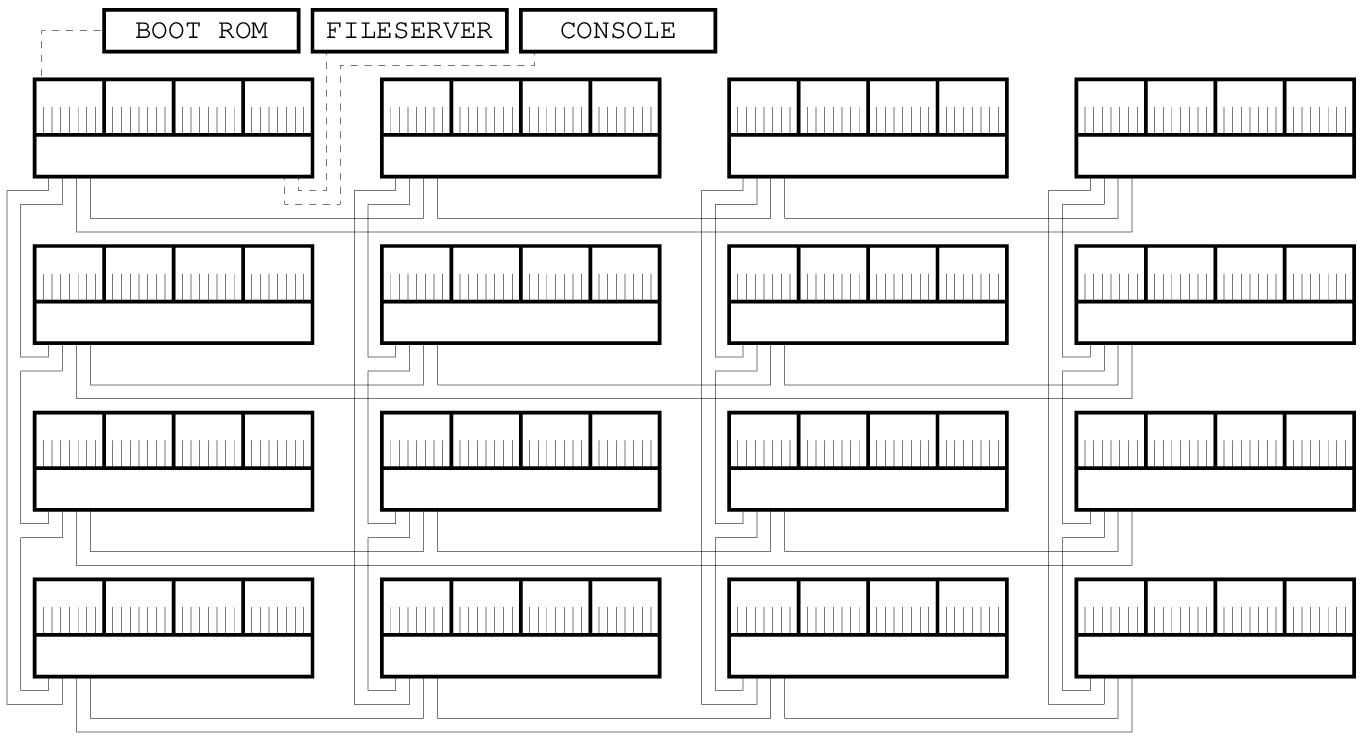}{: 16 processor test setup}).

Only for the first simulator three peripheral lines
are externally connected:
A character stream input line, a corresponding output line,
and a bidirectional connection to an external file server tool
to allow access to files external to the simulation.
Additionally, the first node is configured to accept the
initial boot program via a specialised line
connected to a boot ROM.
\ffig{15}{b}{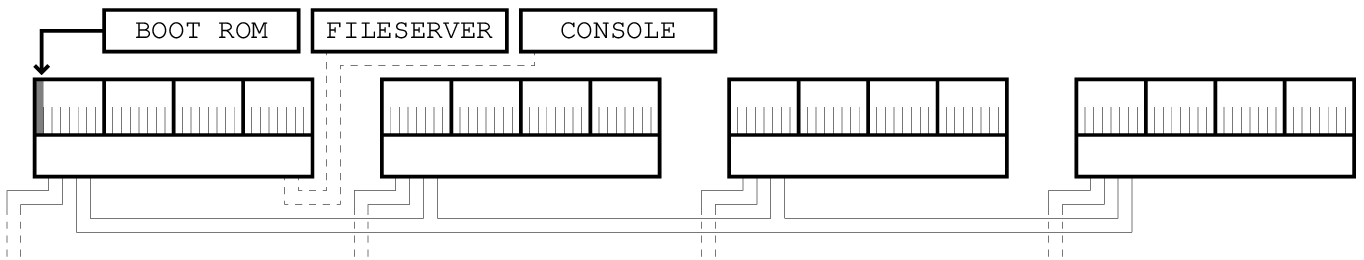}{: initial boot program loaded into first node}

\subsection{Boot Code Supply}

At startup time, each node executes a short first stage
loader code snippet (see \lref{2016os1}).
On the first node,
it reads the operating systems initial boot program from a
simulated boot ROM, i.e. from an external file
(\tfig{15}).
This boot program is composed of code to perform
system initialisation
(boot sequencing, processor enumeration, and routing table setup),
provide peripheral access
(console input and output, file server access),
and runtime process creation support
(process loader, dispatcher, and process schedulers)\footnote{The
prototype implementation is less than 4~kB of executable code.}.

Once loaded into the first node
and started as a single boot process on its first core,
the boot program starts the processor enumeration process
and the console driver. Then it starts four message distribution
processes -- one for each external link --
which are needed to avoid message congestion
(\rfig{16}{htb}{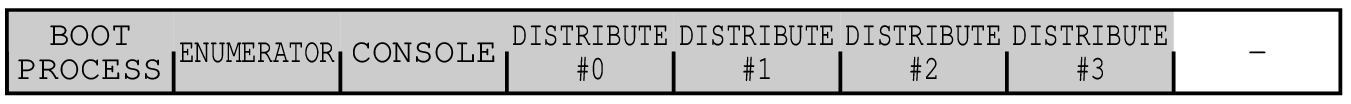}{: first node on first processor during system initialisation}).

The latter is done on
each processor that is started later on, not only on the first node
(\rfig{17}{htb}{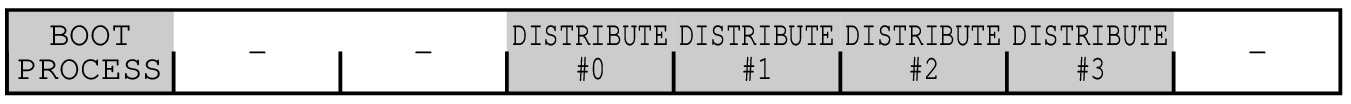}{: first node on other processors during system initialisation}).

\subsection{Boot Code Distribution}

The boot program sends its own code over the
external links, causing it to be accepted by the
first stage loaders of its four neighbouring processors.
It uses an exact copy of its own code with one single
variation to allow the receiving node to detect
that it is not the first node in the system
(\rfig{18}{htb}{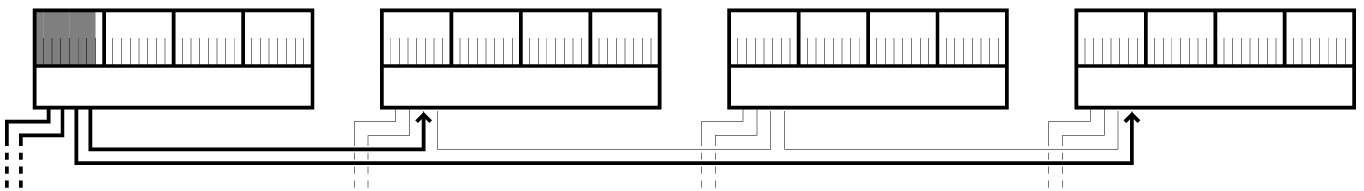}{: boot code distribution, phase 1}).

Not knowing the external links grid topology in advance,
a node may receive initial code more than once,
so it will subsequently discard any extra copies it will receive
(\rfig{19}{htb}{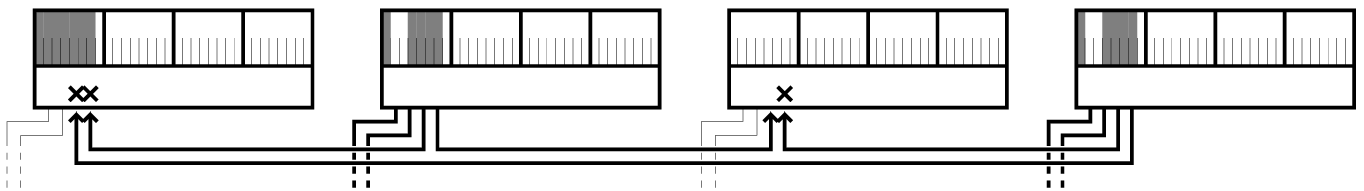}{: boot code is discarded, when received again ($\times$)}).

\label{bootcodedistr}
Later during system initialisation,
just before switching to routine system operation,
the boot program
sends a partial copy of itself
-- only comprising the scheduler process code --
to the remaining three nodes local to the processor
(\tfig{20}).
Both processor enumeration and routing tables
are resources common to all nodes in a processor,
so further initialisation is not needed on
these additional nodes.
\ffig{20}{htb}{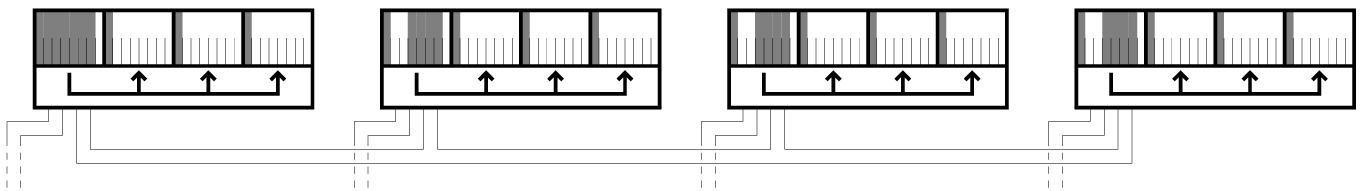}{: scheduler code distribution}

\subsection{Processor Enumeration}

Initially, routing tables are empty for all processors,
and only the first node is aware of its identity.
As the calculation of routing tables requires processors to
be enumerated, the latter is undertaken first.

The processor enumeration process addresses all processors,
one by one, starting at the first processor, causing each
processor to contact its four neighbours -- again one by one --
and assigning numbers to them
(\rfig{21}{htb}{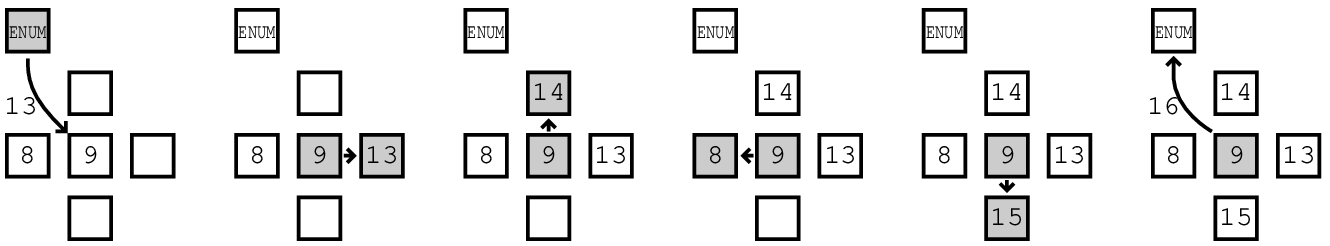}{: processor 9 enumerating its four neighbours}).
Note, that enumeration is done per processor,
because the implicit numbering of the contained nodes and cores is predefined.
This procedure requires the
neighbouring processors to check with the contacting processor,
and make sure they both know on which of the external physical
links they are located with respect to the other processor.

While the contacting processor is able to send a message
explicitely to one of its physical neighbours,
the neighbouring -- receiving -- processor is not aware of the origin
of the message. Therefore it will send a response
to all of its four neighbours in turn, a message including
the original contacting processors number as well as
its own physical link number. Three of these
neighbours will drop the message silently as they
do not feel addressed, whereas the original contacting
processor will detect the valid response, and finally
inform the newly enumerated processor about which
physical link corresponds to their relation.
In doing so, it sets up a route between the two processors
in question
(\rfig{22}{b}{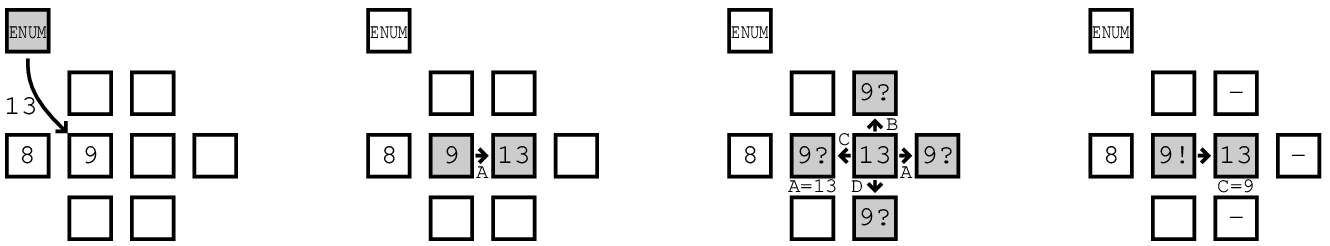}{: processor 9 negotiating the link with its first neighbour}).

Furthermore, as the processor currently in charge of
enumerating its neighbours is either the first processor,
or is reachable via processors that have been enumerated earlier,
it is able to set up a route from the newly
enumerated processor via itself all the way to the first processor.
Simultaneously, the first processor is informed on how
many processors have been assigned new numbers,
so that these processors will subsequently be addressed by the
enumeration process themselves.

The process of enumeration is continued until no more
unenumerated processors are found.
By then, all processors are able to send messages to the
first processor, and the first processor is able to send messages
to all of them, but the routing tables are incomplete
in that they do not allow sending messages between
arbitrary processors.

Obviously, the numbering scheme that will result from
this algorithm is far from optimal. An optimised
algorithm should choose an enumeration to reflect
the physical network topology, so that optimisation of the
resulting routing table, and thus reduction of
routing table size, can be achieved.

\subsection{Routing Tables}

To calculate complete routing tables,
the processor enumeration process addresses all processors again,
one by one, causing them to send some path establishing
message to all four neighbours, who in turn will propagate
these messages to all neighbours, and so on, increasing
the hop count of the message each time it is resent.

All processors keep track about the minimum distance
by inspecting the hop count of the message, and thus
find out the corresponding neighbour and link to
store into the local routing table,
for which the distance to the originating processor is shortest.
The algorithm is similar to the second one presented
by D{\ij}kstra \lref{1959ed}, differing in that it proceeds
asynchronously. This is the reason why each processor
must keep track of the minimum distance, because
it cannot rely on some global algorithm loop count.

Each time a message is found to indicate a lower distance
at some processor, the global enumeration
process is asked to increase its balance by three,
but when instead no lower distance is found,
the balance will be decreased by one.
This way the enumeration process will find the routing
table calculation for one processor to be complete simply
by waiting for the balance to be equalised.
It will address the next processor only when the
previous one has finished its routing path determination,
not because the algorithm asks for it,
but to avoid message congestion and system deadlock.

This algorithm to fill the routing tables offers no
optimisation other than distance calculation. A better
algorithm should at least account for static traffic
optimisation. Dynamic traffic optimisation, e.g. blocked or jammed
route by-passing, would be an advanced system run time task.

\subsection{Routine System Operation}
\label{routinesysop}

As soon as all processors have completely calculated their
routing tables, the system switches from initialisation
mode to routine system operation: The first node shuts down all
processes except the console, then newly starts the
dispatcher process, the loader process, and the file server
(\rfig{23}{htb}{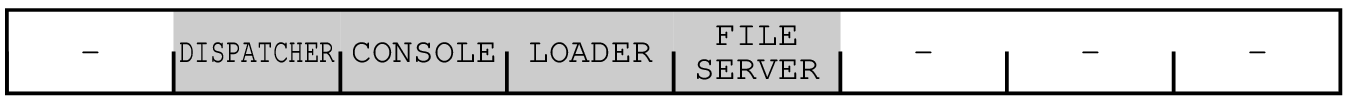}{: first node at routine system operation}).
Finally it loads an arbitrary initial user process
-- named \term{init} -- by means of normal process creation.

All other nodes shut down all processes as well,
but only start one process -- the scheduler
(\rfig{24}{htb}{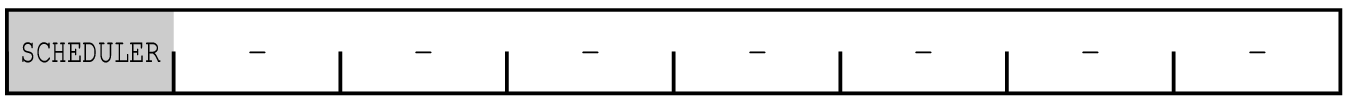}{: any other node entering routine system operation}).
Note that no scheduler is started
on the first node,
because the dispatcher potentially will use up
large amounts of the local memory to keep track of
system wide resources.

Up to this point, only one node per processor was active.
The remaining nodes are now fed with just the schedulers code,
as described earlier (see section \ref{bootcodedistr}).

\subsection{Dispatcher and Schedulers}

With the commencement of routine system operation, a process may no longer
be started on a statically fixed location, but invoking a new
process is subject to system resource management.
The initial user process \term{init}
mentioned above is the first process to follow this rule.

The operating system not only needs to keep track of the
resources, but also provides means to make use of them.
For reasons of simplicity, one single process keeps track
of all nodes and how many cores and how much memory they
currently have at their disposal.
This process -- the dispatcher -- will accept a request
to start a new process, and redirect it to an appropriate
node. As the nodes hardware is not capable of starting
a new process by external request, another node local
system operating process must perform this action instead.
This is the reason why each node runs a seperate instance
of the scheduler process. Additionally, the scheduler process
is responsible for local memory management, and it will
inform the dispatcher process each time the amount of
node local free resources changes: After process creation,
and when a process has stopped. For the latter reason,
it is also the scheduler that will receive process
exception messages informing it about process termination.

With the request to start a new process
the dispatcher process accepts the port identification of the
requesting process, information about the relative amount of memory
the process shall use (the \term{dimension}),
and the binary code stream (see \lref{2016os2}). 
It calculates the absolute amount of memory to allocate for the
process and sends all the information on to a selected scheduler.
It then resets its resource record for the selected node to zero
to avoid allocation conflicts.

The scheduler -- receiving the allocation information and the
binary code stream -- will determine a contiguous memory block for the
code and one for the data, write the code into the first block,
and start process execution.
Then it acknowledges process creation to the requesting process,
so the latter can access the control channel.
Finally the scheduler will recalculate the amount of available memory
and send a report to the dispatcher, together with the number of
available cores.

Clearly, in a system with a large number of cores, to burden
a single dispatcher process with core selection
would make it a bottleneck, so an optimised dispatcher should
be implemented as a set of distributed processes.

\subsection{Loader}

When starting a new process, usually its code is not available
in memory, but resides in an external file, which is known by name.
Same as the dispatcher, the loader process serves the task
of starting a new process, but instead of taking the binary code
as input, it accepts the name of the process, asks the file server process
to load the corresponding external file, and hands the executable
binary code over to the dispatcher
(\rfig{25}{htb}{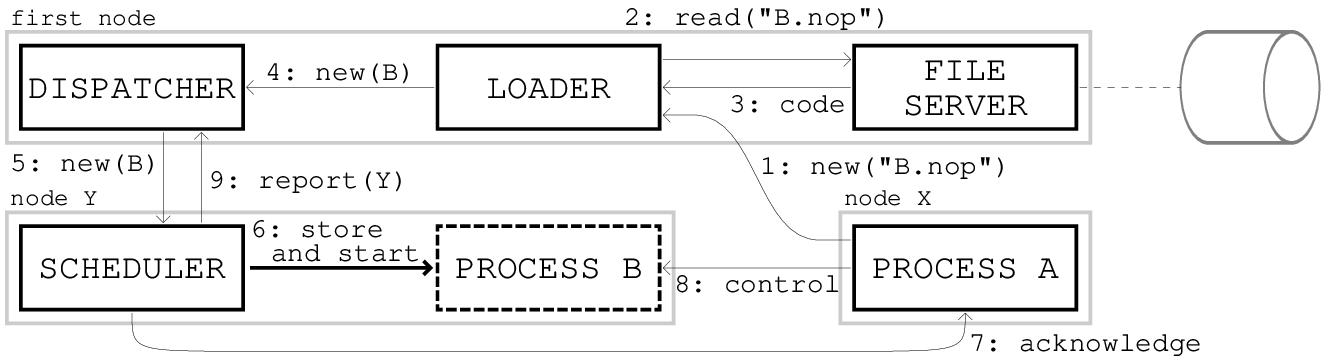}{: process A starting a new process B by name}).

\subsection{File Server}

The file server process mandatorily resides on the first node,
as it must make use of the peripheral line that is bidirectionally
connected to an external service
that provides access to files.
Any process may connect to the first port of the file server
and ask for a file to be read or written.
Files are always read or written entirely, there is no provision
to read partial files, or to change existing files.
Reading or writing a file is done chunkwise, to avoid a single
process blocking other processes to access files simultaneously.
The protocol provided by the file server is similar to
TFTP \lref{1992ks}, cut down to a minimum.

\subsection{Console}

Like the file server, the console process mandatorily resides
on the first node in order to be able to access the two peripheral lines
that constitute standard input and output of the system.
The console process provides two ports, the first one for input,
the second for output.
Note, that data transmission on both ports is not byte oriented,
but word wise, as is all data processing in the
\term{Null Operand Parallel} processor system.
For text message encoding,
usually Unicode \lref{2009uc} is used directly,
and it is up to the invoking entity to convert data from and to
the system to the character set that is handled by its terminal.

Any process may connect to the first console port to exclusively
reserve input.
It sends one word indicating its own port, to inform the console
about where to send console input, but it must not end the connection,
unless it intends to release the console input.
Data sent to the second port will be transmitted to the output
peripheral line.

\subsection{User Processes}

When routine system operation is entered,
the last task performed by the boot process is to load
an initial user process named \term{init}.
In a simple system configuration \term{init} will refer to
a shell to allow a user to directly control the system.
In a more mature system design a separate initialisation
program may launch various system services, but for test
purposes a shell will do.

User processes may handle both input and output channels,
the number of which may differ depending on the purpose
of the process. E.g., a process designed to merge two streams
would provide two input ports and one output port,
while a process that is to duplicate a stream
would provide one input port and two output ports.

It is up to the user process to announce the number of
ports it provides.
For the output ports,
it is informed by the invoking process about the destination ports,
and thus is able to connect to them autonomously.
For the input ports it provides,
it needs to inform the invoking process about these ports,
and subsequently wait for incoming connections,
because it is always the sending process that initiates
the channel connection.

This negotiation is done on the control channel of the user process,
and by convention, all users processes follow the same scheme
(\rfig{26}{htb}{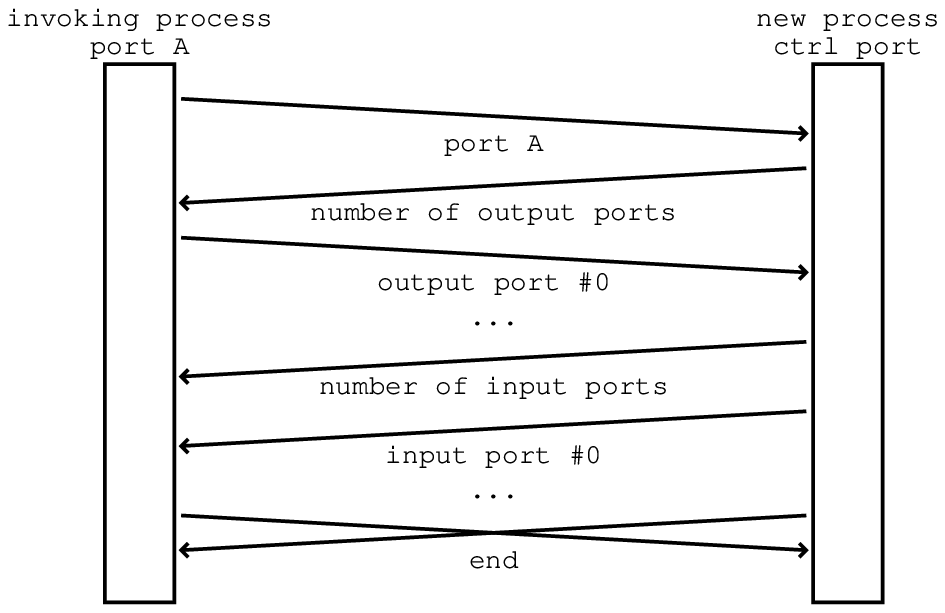}{: initial port negotiation for a user process}).

\subsection{User Shell}

The user shell process is a simple text based command line interpreter
(see page \pageref{sourceusershell}).
It scans an input line for tokens, invokes the appropriate user processes,
and makes them connect to the right input and output ports.

In a higly parallel channel based environment, a set of processes
does not simply work on files, or in a linear pipeline,
but needs to be connected in more complex networks.
To avoid a notation for explicit port addressing, an approach is chosen
similar to the \term{Polish Notation} --
initially introduced in the 1920s by Łukasiewicz
for the logic of propositions \lref{1970jl}.
No specific operator symbols, parantheses, or channel numbering are needed.
Setting up a network of user processes is done by
writing their names in the desired order.
A stack of port identifiers is maintained internally to
keep track of the order in which to connect user processes.

Each command line is evaluated according to the following syntax:

\vspace*{2ex}

\noindent
$line ::=$ \{ $term$ \}\\
$term ::=$ \{ $command$ $\vert$ $string$ \}\\
$command ::=$ $word$ $\lbrack$ ``{\tt :}'' $number$ $\rbrack$\\
$string ::=$ ``{\tt "}'' \{ $char$ \} ``{\tt "}'' \{ ``{\tt "}'' \{ $char$ \} ``{\tt "}'' \}\\

The line is evaluated from left to right.
For each command, the corresponding process is started.
During port negotiation with a single command process,
the shell takes port identifiers from its internal stack one by one
to provide the process with output ports as requested.
Then, it accepts input port identifiers from the process,
and pushes them onto the stack one by one.

At the beginning of a line, the stack is conceptually initialised to
hold an infinite amount of identifiers for the second port of the
console process, i.e. standard output.
This way, any open output port will eventually be directed
to the system console.

At the end of a line, supposed the stack is not empty
-- not counting the virtually infinite amount of console output ports --
the top most port is connected to the console input,
but any further ports remaining on the stack are immediatly
fed with an end token\footnote{This
is equivalent to input from {\tt /dev/null} with Unix.}.

Regarding the notation, the processes in a command line are
invoked from left to right, but data flow usually will be
from right to left, because the shell will provide port
identifiers from commands to the left as output ports,
and pass input port identifiers on to commands to the right.

Immediate text data may be injected into an input port
of a process by writing it in double quotation marks.
The shell will transmit the text data into the corresponding
input port of a command to the left promptly.

To start a process with a given \term{dimension}
(see \lref{2016os2} for details),
the command may be followed by a colon and a number,

The stack order is fixed, so currently changing it would need
an extra permutation process. Alternatively, a special notation
could be introduced for direct stack permutation within the shell.

\ffig{27}{b}{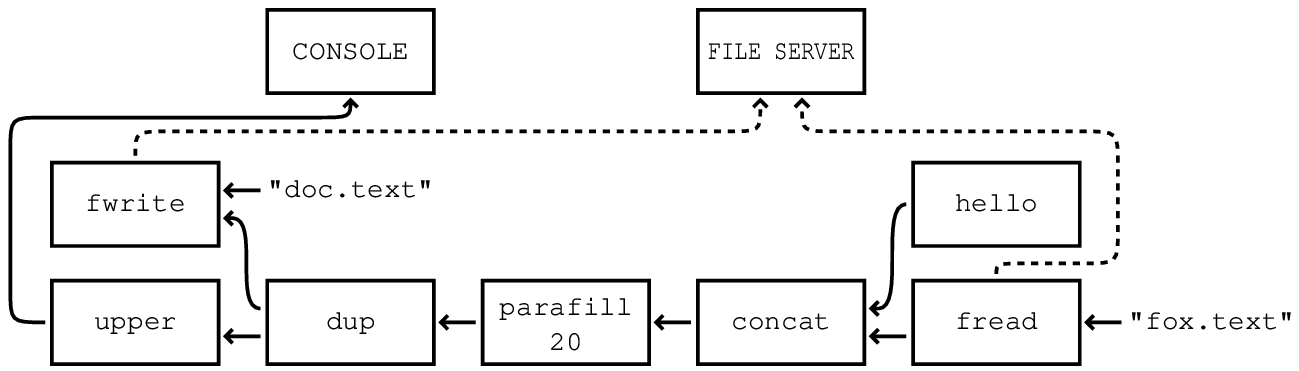}{: data flow in a user processes network}
The following example reads a text from a file {\tt fox.text},
concatenates it to the output of a \term{Hello World} program,
formats it to keep a line width of at most 20 characters,
then duplicates the result, storing one copy into
a file named {\tt doc.text}, and converting the other copy
to upper case, finally sending the result to console output
(\tfig{27}).\newline
\newsavebox\foxex
\begin{lrbox}{\foxex}\begin{minipage}{\textwidth}
\begin{verbatim}


> upper fwrite "doc.text" dup parafill:20 concat hello fread "fox.text"
HELLO WORLD THE
QUICK BROWN FOX
JUMPS OVER THE LAZY
DOG.
\end{verbatim}
\end{minipage}\end{lrbox}
\resizebox{0.89\textwidth}{!}{\usebox\foxex}

\subsection{Application Software}

Eventually, the purpose of a computer is to be applied,
so what is needed is application software. There is a vast
amount of possible applications, which is not addressed
here in detail at all. However, a minimum set of tools
is necessary to check system functionality and to support
basic data processing. Furthermore, a few representative
yet simple examples are given to show applicability of
the system designed.

Two tools serve system inspection. 
Both the dispatcher and the schedulers allow their state
to be queried, so the tools only need to send a query to the
system process and output the formatted result to the console.
The first tool, {\tt qdisp}, provides a table on the dispatchers
state, e.g. for a four processor system:

\begin{verbatim}
> qdisp
 8a: 0/0000/0000
 8b: 0/0000/0000
 8c: 7/3eff/0000
 8d: 7/3eff/0000
 9a: 7/3eff/0000
 9b: 7/3eff/0000
 9c: 7/3eff/0000
 9d: 7/3eff/0000
10a: 7/3eff/0000
10b: 7/3eff/0000
10c: 7/3eff/0000
10d: 7/3eff/0000
11a: 7/3eff/0000
11b: 7/3eff/0000
11c: 7/3eff/0000
11d: 7/3eff/0000
\end{verbatim}

Each row displays the resources of one node, with four nodes
{\tt a} through {\tt d} per processor, which in turn are numbered from {\tt 8} upwards.
The first value is the number of available cores, the
second and third are the size of the largest and the second
largest contiguous memory area at the node. In the example, the
first node {\tt 8a} is not available, as it is used for system
purposes (see \ref{routinesysop}), the second node {\tt 8b} is
blocked by its scheduler during startup of a process -- which
in this case is the inspection tool itself. All the other nodes
are currently idle and empty.

The second tool, {\tt qsched}, shows a schedulers state,
by default the one on the node it is executing on, but by providing
a dimension $4{\times}processor+node$ -- starting at {\tt 33}, because
processors deliberately are counted from {\tt 8}, and the first node on the
first processor does not provide a scheduler --
a different node may be selected:

\begin{verbatim}
> qsched:33
 8b: 0101..4000 t:9162d70d c:00029a43
1: 0101..01f0/01f1..03c7
2: 03c8..0457/0458..057a
3:
4:
5:
6:
7:
\end{verbatim}

The first line indicates the nodes identity {\tt 8b},
the memory range available for user process allocation,
the nodes current time counter, and the total instruction
cycle counter for the node. The subsequent table provides
for each core that currently executes a process the constant pool range
and the data pool range. Only seven cores are listed,
as core number {\tt 0} is used by the scheduler. In the
example, the first core executes the {\tt init} process,
which is the shell, and the second core executes the
{\tt qsched} tool.

Furthermore, a set of simple tools is available for basic
data processing:

\begin{description}
\item {\tt fread} -- reads a file name on the input channel,
  connects to the file server to read the file,
  and write the contents to the output channel.
\item {\tt fwrite} -- reads a file name on the first input channel,
  reads data from the second input channel, and
  connects to the file server to create and write the file.
\item {\tt concat} -- reads data from the first input channel and writes
  it to the output channel, then reads data from the second input
  channel and writes it to the output channel.
\item {\tt dup} -- reads data from the input channel, and writes
  it to both of its two output channels.
\item {\tt buf}:$n$ -- is given a dimension, and maintains a ring buffer
  of that size, reading data from the input channel to the buffer,
  and writing data to the output channel from the buffer. When the
  buffer is full, it does not spill input data, but blocks instead --
  until the output channel is accepting data again and
  consequently the buffer will no longer be full.
\item {\tt merge} -- reads data from two input channels,
  message-wise
  in random order according to availability, and writes all data
  to the output channel.
\item {\tt nil} -- writes an end token to the output channel. This
  is equivalent to input from {\tt /dev/null} with Unix.
\item {\tt absorb} -- reads data from the input channel and discards it.
  This is equivalent to output to {\tt /dev/null} with Unix.
\item {\tt hello} -- write a message to the output channel.
\item {\tt upper} -- reads data from the input channel, and writes it to
  the output channel, converting latin lower case letters to upper case
  (see page \pageref{sourceuppercase}).
\item {\tt parafill}:$n$ -- is given a dimension, the designated line length.
  Reads text from the input channel, and writes it to the output channel,
  reformatting blocks of text to limit and fill line length to
  the dimension, whenever possible.
\end{description}

Process instantiation may be different, depending on the needs.
The system inspection and data processing tools listed above
are loaded from an external file server and executed individually
as requested, and they terminate when their work is done.
This is comparable to system tools invoked from shell or script
with Unix.

Other processes will be started once and never stop, providing a service.
For utilisation, another process may connect to the control channel,
provide configuration, write and read data, and so forth, thus
occupying the service for a while. The difference is that freeing the
service does not stop the process, but makes it available to the next
user. This type of process occassionally is called a \term{daemon} --
or a \term{driver}, when it provides access to hardware functionality
such as peripherals. The {\tt dispatcher} is an example for a \term{daemon},
while {\tt file server} and {\tt console} are examples for drivers.

Large programs will be implemented as a set of processes, invoking
each other. It is a design question whether parts of the program will
run permanently, as a daemon does, or be started anew whenever its
functionality is needed. The latter is similar to a function called,
but the difference is that starting the process anew involves loading
the code to memory again. There is not much use in keeping the code
around in memory elsewhere, as it does not make much of a difference
whether the code is transferred from one memory location to another
and then started, or it is kept running as a daemon.

When it comes to implementing large applications, a variety of
strategies will be used to spread its functionality to processes
as required. Two characteristic options are to replicate the key
algorithm in an application and split up the data accordingly,
to increase throughput, or to split up a complex algorithm
into parts, see table 28.
The first option is popular e.g. in image processing.
The latter option may prove useful where processing
of a data stream is time consuming, and it is possible to split
up the algorithm into stages of a pipeline (multiple passes).
Moreover, it is mandatory where the algorithm is to complex to fit
the code into the local memory of a single processing unit.
A problem solving algorithm may also be split into processes
to investigate different solutions simultaneously (see \lref{1994os}, p.67$f$).

\begin{table}[htb]
\begin{center}
\begin{tabular}{|p{1.8in}|p{1.5in}|p{1.6in}|}
\hline
& simultaneous & sequential \\ \hline
homogenous \newline (an algorithm replicated) & data split into chunks or tiles & multiple identical passes \\ \hline
heterogenous \newline (an algorithm split) & partial algorithms, possibly interacting & pipelined processing of data \\ \hline
\end{tabular}
\end{center}
\begin{center}
{\it table 28: Multiple Process Algorithms}
\end{center}
\end{table}

Implementation of a basic set of development tools
is not required initially, but will be needed
later (e.g. editor and compilers).

\section{Results and Discussion}

Combining the hardware sketch with a comprehensive dynamic
operating system allows the operation of general purpose application
software comparable to what contemporary work stations allow.
It is obvious that approaches to software design will differ
substantially from the usual practice with shared-memory
based systems. Software existing for the latter cannot be
ported just as is, but it has to be rewritten, and in many
cases it has to be completely rethought and redesigned:
\cite{The effective use of concurrency requires new algorithms
designed to exploit this locality} (see \lref{1987ms}, p.36).

Based on the assumption that up to 6000 cores fit onto a single chip,
the system clock frequency at 400~MHz, shared among every eight cores,
so each core is executing instructions at 50~MHz,
a rough estimation for the performance could be 300000 MIPS per chip.
Though this is just a theoretical upper bound,
the design strongly suggests the presumption
that performance per die area will be much higher than with
existing shared-memory based systems, by the fact alone
that the percentage of silicon lying dead and unused
is much lower, as Hillis anticipated (see \lref{1985dh}, p.4).

For the sake of being able to achieve a first draft,
lots of details have been simplified, and it is beyond question
that a really usable system requires refinement in most
respects and research in various fields.
E.g., the instruction to start a new process takes five
operands from the stack, which makes it difficult to
implement in hardware.

\subsection{Instruction Set Architecture}

A zero address stack design has been chosen for three reasons:
First, implementation of a simulator for it is quite simple,
and second, code generation is straightforward for a compiler
designed according to \lref{1977nw}. Third, instruction
encoding is compact compared to other architectures, with
only eight bits for a full instruction. However, functionality
per instruction is quite restricted, e.g. an extra instruction is
needed for each operand that needs to be fetched from memory.
To increase performance, the stack oriented design may be
optimised (see e.g. \lref{2015jb}), or
another instruction set architecture may be chosen.

\subsection{Hardware Supported Scheduling}
\label{hwsupportsched}

With the current design,
in each node, one core is used for the scheduler, to accept
and start a new process, to detect its termination, and to
report current allocation to the dispatcher.
Adding associative circuitry (see \lref{1996hv}, p.53) to manage memory allocation,
it is possible to replace the software implemented scheduler
by supportive hardware,
see \rfig{29}{htb}{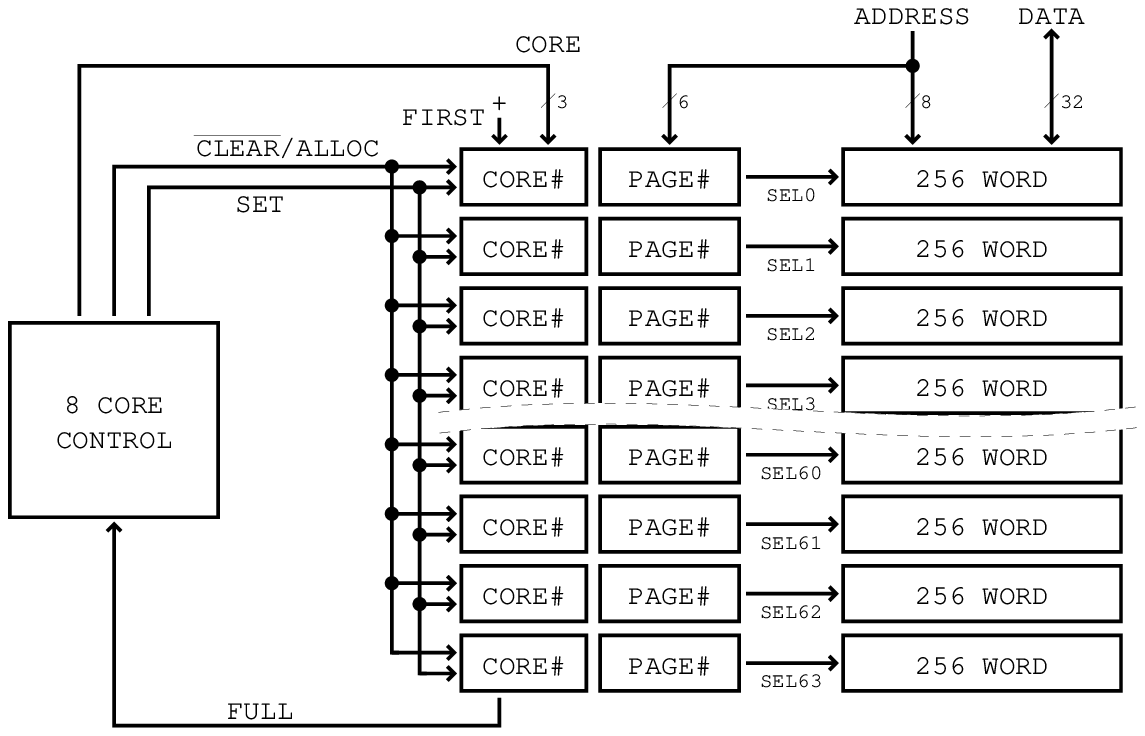}{: hardware scheduler 64 page allocator overview},
and \rfig{30}{htb}{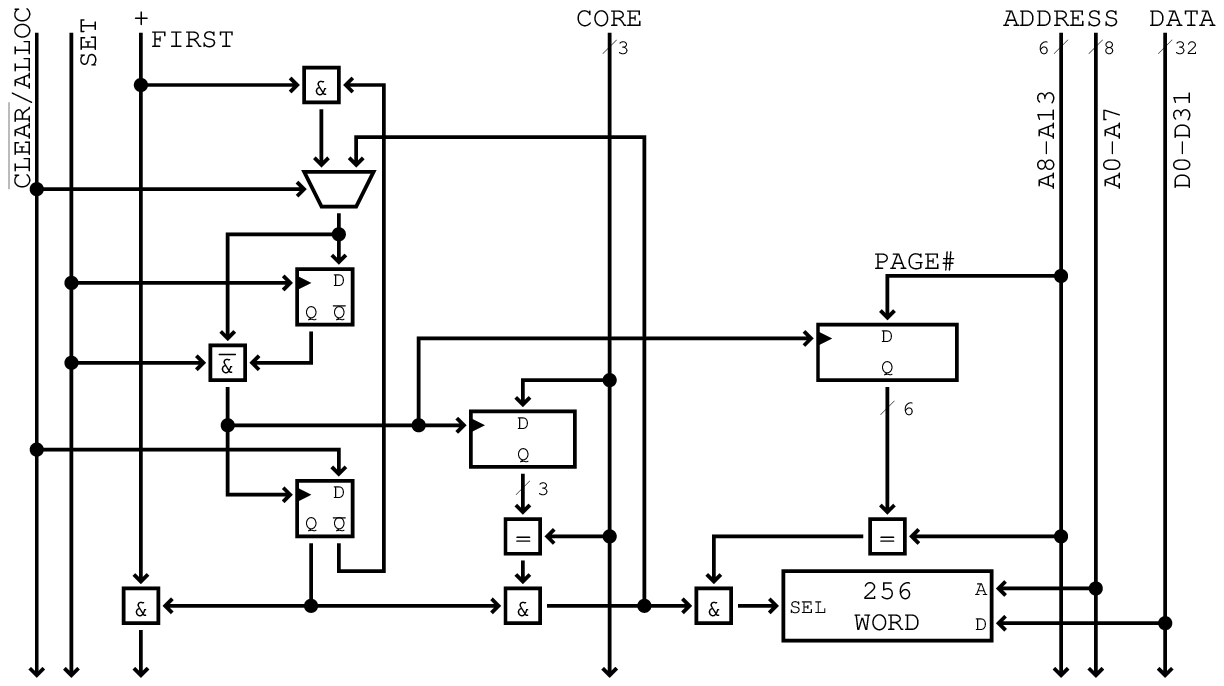}{: hardware scheduler allocator single page detail}
\footnote{The depicted circuitry allows
allocation of a page by applying the desired core and page number,
set {\tt ALLOC} high, and then signal a high pulse on {\tt SET}.
Deallocation is done by applying the desired core number,
set {\tt ALLOC} low, and then signal a high pulse on {\tt SET}.}.
The dispatcher then would send a process to start simply to
some special port on the destination node. Both upon start
and termination the core hardware would automatically send
a resource indication message back to the dispatcher.

\subsection{Channel Virtual Addressing}

Like global memory is the main spatial resource on shared-memory
systems, processing units are the main resource on the proposed design.
When resources are exhausted, idle processes might be frozen,
and their current state and memory content swapped out to some
external memory. To be able to restore and continue execution of
such a process, \term{channel virtual addressing} \lref{2013os}
is needed, because
no guarantee can be given for the process to be loaded back to
the identical location where it resided earlier.

Other than for memory as a resource, which is flat and thus usually
evenly accessible in whichever order it is arranged on a single node,
access to channels
depends on the network topology, and congestion may occur when
too many messages are sent over a single route with no alternative
available. Here, \term{channel virtual addressing} will also be useful
when processes shall be relocated for traffic optimisation.

\subsection{Efficient Message Routing}

Efficient routing support is not trivial and needs to be addressed
separately.
It is subject to research for a long time -- both algorithmic basics \lref{1959ed}
and theoretical background \lref{1987pu} --
and extensively (e.g. \lref{2006pf}),
and still is extensively being explored (e.g. \lref{2009mf}).
As with the work presented, algorithms should address
general network topologies, initially unknown,
and with the nodes initially unlabelled.

\subsection{Endianness and Alignment}

Direct byte addressing had been chosen,
among others, to simplify text processing at a time when memory
was expensive, and text was encoded at no more than some eight bit
per character.
Byte adressing causes memory addresses to be counted in bytes,
while memory access is word-wise, today usually at 32 or 64 bits per word.
Access to odd memory addresses causes data bus misalignment,
which is handled either by shift adjusting data at handover
from memory to registers, or by trapping.
Once available in a register, the order in which contiguous bytes
are located with respect to each other depends upon \term{endianness}
of the system. As endianness differs on various systems, it is
an everlasting wealth of confusion when it comes to transferring
data or porting software from one system to another.

To allow non-latin text encoding, and to avoid trouble through text recoding,
today text characters usually are encoded with more than eight bits.
The Unicode Standard \lref{2009uc} is the most universal approach,
with characters represented by 21 bits each. The simplest encoding on
a system with 32 bits per word is to store one character per word.
The Unicode Standard refers to this representation as UTF-32.
Popular representations like UTF-8 or UTF-16 are used to save space
in data storage, but they can be taken for specific data compression
formats. As such, where availability of storage is an issue,
explicitely compressing text to UTF-8 may well be a solution.

Both endianness confusion and misalignment annoyance
vanish when byte addressing is given up (see \lref{2016os1}).

\subsection{Data Transmission Protocols}

Peripheral drivers need to handle incoming and outgoing data
streams in real time. The structure of the data is defined
by the respective protocol standard, often specifying various
interdependencies among part of a data stream, e.g. a checksum
to depend on a subsection of the stream.

The receiver will have to evaluate such interdependencies,
and the sender will have to generate data to fulfil them.
A protocol may or may not be designed to allow streamed
processing -- like it is also used in cut-through switching --
i.e. fulfilling all interdependencies with no
more than a limited fixed amount of local memory.
TCP \lref{1981jp} e.g. transmits a checksum of the data that follows,
so potentially a full data packet has to be stored temporarily,
before the checksum can be calculated and the header sent,
eventually followed by the data.

\subsection{Memory Hierarchy}

The single node von Neumann computer stores data in registers and memory.
Magnetic drums and tapes, harddisk and floppy disks are added to
provide permanent data storage. Cache memory has been introduced
to compensate for slow DRAM access in relation to processor
execution speed, and for the memory access bottleneck in a
shared-memory system.

Replacing global DRAM by local fast memory renders cache memory
superfluous. Large amounts of global memory are either superseded
by large amounts of processing units, or by external memory
resources, be it e.g. harddisk storage or DRAM based offline storage.
The latter may be implemented by attaching DRAM to selected nodes.
To simplify system application, all these external storage devices
shall be controlled by a single universal type of interface.
It is subject to further research whether driver processes are
needed to provide access to the single storage devices,
or whether there are more flexible approaches.
Generally, it is recommended to keep the memory hierarchy as
flat as possible and reduce the number of concepts.

\subsection{Floating Point Support}

Floating point arithmetic has been implemented in some of the
first computers ever \lref{1936kz}.
Undoubtly, it is very useful for various categories of applications.
On the other hand, the hardware demand to implement it is higher
than with fixed point arithmetic (see e.g. \lref{1989dm}).
However, increasing the hardware demand for each single core would
foil efforts to integrate as many as possible cores on a given
die area.
Further research is needed to investigate the question how
floating point arithmetic can be supplied with as little
hardware demand as possible.
Possibly the trick is to split up calculation into simple
and fast supportive instructions, and maybe just three or four
of these would do:

\begin{description}
\item $\bullet$ load floating point number into two registers {\tt E} and {\tt M}
\item $\bullet$ store two registers as a floating point number to memory
\item $\bullet$ normalise two numbers with respect to each other
\item $\bullet$ and maybe some shifted register addition instruction
\end{description}

\subsection{Loading and Execution of Processes}

Loading the code for a process usually takes time, the larger
the process to run the longer. Once loaded, starting the
process is constant effort, mostly neglectable.
For applications that need to execute large numbers of processes,
the overhead of loading and starting them can be substantial \lref{2011hh}.

Research is required to find, whether it is efficient to provide
functionality by a daemon instead of loading and executing
a process over and over again, or whether there are ways to
avoid reloading process code, possibly by some method of caching it
instead of just freeing its memory region.

\subsection{Privilege Control}

On shared-memory systems, privilege control is generally achieved
by running lower priority processes with a modified instruction set
to block privileged operations, and by blocking access to regions
of physical memory that do not belong to the process.
The latter is not relevant on a system with no shared memory,
instead access to other processes and their data is only possible
using channel communication. Consequently, privilege control
would ask for restrictions in sending messages to specific ports.
This may be achieved by marking each port with its level of
access permissions, and then blocking incoming messages that
originate from any port with lower privileges. To be able to detect
the privileges of the sender, these would be part of the message
header, as it is automatically composed at the senders outgoing port.

All this can be implemented purely in hardware -- except setting
the acceptance permissions for each single port, which is under
decision of the privileged process.

When more than one user shall use the system and each user is to be
restricted to its own group of processes, some kind of owner
information would need to be added to each message header.
Evaluation of the access restrictions at the receiver of a
message then would be too complex to be implemented in hardware,
but would require supportive software. Reading incoming
data, identification and access permissions of the sender would
be made available by the port hardware, so the receiver can
decide on whether to accept the message or not.

\subsection{Channel Congestion Avoidance}

Due to the limited number of connecting channels from one
node to another -- especially where nodes are located
on separate chips -- it may happen that ongoing and unpaused
transmissions between these nodes make use of all
transmission lines available.
When in this situation an additional connection between
these nodes is required, while completion of the other
transmissions depends on the additional connection to
be established, the system faces deadlock.
The obvious solution should be to fix the algorithm,
so that unpaused transmissions do never depend on
additional channels to be set up.

However, it is subject to further research to find whether
this is always possible, or, if not so, whether it
is necessary to avoid the deadlock by automatically
pausing one of the ongoing transmissions to temporarily
free a connecting channel.
Some processes will be able to decide precisely when
to pause the channel, because they have enough information
on the expected timing of their output data.
Other processes, namely those that receive a stream
and propagate the processed data in turn,
will not be able to figure out timing information.

A heuristic approach is to pause the channel after some
timeout. Depending on the overall application design,
this may work, but it may also cause the deadlock to
be inverted. Another approach is to detect the
incoming data being paused and pause the output stream
accordingly. This latter solution would involve
a simple extension to the hardware functionality for
the detection of the \term{pause} token at the receiver.

\section{Conclusion}

Though much higher performance per die area could be achieved with the
proposed system architecture, most of existing software is inherently
non-parallel, designed for sequential execution on single node
shared-memory systems. Memory allocation and process interoperation
of the two approaches do not match at all, thus most software will simply
not be portable, and consequently needs to be rewritten.

On the other hand, without software the best hardware will be of no use.
Accumulated cost for a contemporary operating system kernel alone is given
to be in the range of hundreds of years,
suggesting that reaching the state of
usability, which we are used to expect from a mature system,
is quite high. This may be a reason why implementation of multinode
systems has been avoided for long -- but there is no way around:
Rewriting all the software is expensive.

However, research and prototype implementation 
are to prove practicability of the basic design principles,
they do not need to
fulfill quantitative requirements in the first place.

As has been shown, it is and will be crucial to reduce complexity,
to keep the design as simple as possible. Otherwise the highest possible
number of cores per chip, and thus the highest possible
performance per die area, will not be achieved.

Moreover, it has been shown that both co-design of hardware and software,
and a conceptually simple operating system are doable.
Such an approach may serve as a remedy for the memory bottleneck impasse.

Further research needs to focus on two topics: First, a simplified and more
implementation-oriented design of a hardware prototype needs to be designed.
To be able to realistically simulate parallelism, it should be implemented
in hardware, e.g. FPGA based.

Second, basic algorithms and applications need to be designed and implemented
to demonstrate use cases for the system. Both system related algorithms --
e.g. for message routing -- and application related algorithms --
e.g. for storage allocation -- are equally important for advanced deployment
as well as the determination of more realistic numbers and values concerning
the performance one can expect to attain.

\newpage

\subsection*{Appendix: Source Code Examples}
\addcontentsline{toc}{section}{Appendix: Source Code Examples}

\subsubsection*{Upper Case ({\tt upper.gustl})}
\label{sourceuppercase}

\begin{small}
\begin{verbatim}
process upper(ctrl)
state data, break
port in, out
word i

start
  if not ctrl ? i then           { port negotiation, see figure 26 }
    next break
  done
  ctrl := i                          { respond to invoking process }
  ctrl ! 1                                { number of output ports }
  if not ctrl ? i then        { receive output port identification }
    next break
  done
  out := i
  ctrl ! 1                                 { number of input ports }
  ctrl ! in                    { provide input port identification }
  ctrl ! end                               { port negotiation done }
  next data

on data \ in ? i:                 { read input word when available }
  if (i >= 'a') and (i <= 'z') then
    i := i + ('A' - 'a')
  done
  out ! i               { send processed word to output, may block }
  next data

on data \ in ? end:         { when end of input stream is detected }
  out ! end                     { ... close output stream likewise }
  next break

on break:
stop






\end{verbatim}
\end{small}

\subsubsection*{User Shell ({\tt ulsh.gustl})}
\label{sourceusershell}

\begin{small}
\begin{verbatim}
process ulsh(ctrl)
state prompt, space, error, input, name, dimension, command,
  console, string, quote, break
port in, out, cmd
word n, c, d, i, p
word buf[80]
word tos
word stack[128]

procedure nilinput()
do
  while tos repeat #stack times
    tos := tos - 1                  { remove all ports from stack, }
    cmd := stack[tos]               { ... closing each single port }
    cmd ! end
  done
return

start
  if not ctrl ? n then    { port negotiation, when ulsh is invoked }
    next break
  done
  ctrl := n                          { respond to invoking process }
  ctrl ! 1                                { number of output ports }
  if not ctrl ? p then        { receive output port identification }
    next break                  { for ulsh, output and input would }
  done                   { ... usually be connected to the console }
  out := p
  ctrl ! 1                                 { number of input ports }
  ctrl ! in                    { provide input port identification }
  ctrl ! end                               { port negotiation done }
  tos := 0          { stack of port identifiers initially is empty }
  next prompt

on space:
  if c = 10 then                         { if upon end of line ... }
    if tos then                { the port stack is not empty, then }
      tos := tos - 1                       { ... take the top port }
      n := tos
      nilinput()
      cmd := stack[n]
      next console       { ... and read input from console into it }
    done
    next prompt
  done
  next input

on console \ in ? c:
  cmd ! c
  if c = 10 then
    cmd ! end
    next space
  done
  next console

on error:
  out ! 'f'
  out ! 'a'
  out ! 'u'
  out ! 'l'
  out ! 't'
  writeln(out)
  nilinput()
  next prompt

on prompt:
  out ! '>'                            { print some command prompt }
  out ! ' '
  out ! end
  next input

on input \ in ? c:
  if c = '"' then                { some quoted string, see page 40 }
    n := 0
    if tos then                 { ... directly fed into next port, }
      tos := tos - 1                 { ... when stack is not empty }
      cmd := stack[tos]
    else
      cmd := p        { otherwise sent directly to standard output }
    done
    next string
  elseif c > ' ' then
    buf[0] := c
    n := 1
    next name
  done
  next space

on name \ in ? c:
  d := 0
  if c = ':' then           { command name followed by a dimension }
    next dimension
  elseif c > ' ' then
    if n = #buf then                 { command name buffer overrun }
      next error
    done
    buf[n] := c              { collect characters into name buffer }
    n := n + 1
    next name
  done
  next command

on dimension \ in ? c:
  if c <= ' ' then
    next command
  elseif (c < '0') or (c > '9') then
    next error
  done
  d := d * 10 + c - '0'
  next dimension

on command:
  cmd := PORT_LOADER           { connect to system loader process, }
  cmd ! cmd                            { ... see figure 25, step 1 }
  cmd ! d                                  { send dimension first, }
  i := 0
  repeat n times
    cmd ! buf[i]               { ... followed by command file name }
    i := i + 1
  done
  cmd ! '.'
  cmd ! 'n'
  cmd ! 'o'
  cmd ! 'p'
  cmd ! end
  if not cmd ? i then            { receive schedulers acknowledge, }
    next error                         { ... see figure 25, step 7 }
  done
  if cmd ? end then done
  if not i then
    next error
  done
  cmd := i        { connect to new process, see figure 25, step 8, }
  cmd ! cmd      { ... and provide it with own port identification }
  if not cmd ? n then   { wait for desired number of output ports, }
    next error                                 { ... see figure 26 }
  done
  repeat n times         { provide new process with n output ports }
    if tos then                      { ... if available from stack }
      tos := tos - 1
      cmd ! stack[tos]
    else
      cmd ! p        { otherwise let it connect to standard output }
    done
  done
  if not cmd ? n then             { wait for number of input ports }
    next error
  done
  repeat n times            { receive n input port identifications }
    if not cmd ? i then
      next error
    done
    if tos <= (#stack - 1) then { ... and push them onto the stack }
      stack[tos] := i        { ... to be connected to output ports }
      tos := tos + 1                    { ... of further processes }
    done
  done
  if cmd ? end then done        { expected end of port negotiation }
  cmd ! end
  if tos > (#stack - 1) then                 { port stack overflow }
    next error
  done
  next space

on string \ in ? c:
  if c = '"' then
    next quote
  done
  cmd ! c
  next string

on quote \ in ? c:
  if c = '"' then
    cmd ! '"'
    next string
  done
  cmd ! end
  next space

on input, name, dimension, console, string, quote \ in ? end:
  out ! 's'
  out ! 't'
  out ! 'o'
  out ! 'p'
  writeln(out)
  next break

on break:
stop
\end{verbatim}
\end{small}

\newpage

\subsection*{Appendix: Acronyms}
\addcontentsline{toc}{section}{Appendix: Acronyms}

\begin{longtable}{p{1in}l}
CMOS & Complementary Metal-Oxide Semiconductor \\
CPU & Central Processing Unit \\
CSP & communicating sequential processes \\
DMA & Direct Memory Access \\
DRAM & Dynamic RAM \\
DSM & Distributed Shared Memory \\
DSP & Digital Signal Processor \\
EUV & Extreme Ultra Violet \\
FPGA & Field Programmable Gate Array \\
FPU & Floating Point Unit \\
GPIO & General Purpose Input Output \\
GPU & Graphics Proocessing Unit \\
I/O & Input and Output \\
IP & Internet Protocol \\
MIMD & Multiple-Instruction stream Multiple-Data stream \\
MIPS & Million Instructions Per Second \\
MMU & Memory Management Unit \\
$\mu$P & Micro Processor \\
MPU & Memory Protection Unit \\
NoRMA & No Remote Memory Access \\
NUMA & Non-Uniform Memory Access \\
PHY & Physical Layer Circuitry \\
RAM & Random Access Memory \\
ROM & Read-Only Memory \\
SIMD & Single-Instruction stream Multiple-Data stream \\
SISD & Single-Instruction stream Single-Data stream \\
SMP & Symmetric Multi Processing \\
SRAM & Static RAM \\
TCP & Transmission Control Protocol \\
TFTP & Trivial File Transfer Protocol \\
UART & Universal Asynchronous Receiver Transmitter \\
UDP & User Datagram Protocol \\
UTF & Unicode Transformation Format \\
VLIW & Very Long Instruction Word \\
VLSI & Very Large Scale Integration \\
\end{longtable}

\newpage

\subsection*{Literature}
\addcontentsline{toc}{section}{Literature}

\def\Lit#1#2{\item {\tt [#1]}
#2}
\begin{list}{}{
  \setlength{\labelwidth}{0mm}
  \setlength{\itemsep}{0ex plus0.2ex}
  \setlength{\leftmargin}{6ex}
  \setlength{\itemindent}{-6ex}
  \setlength{\labelsep}{0mm}}
\def\rtit#1{{\it ``#1''}}

\Lit{1866wj}{
  William Stanley Jevons:
  \rtit{The Coal Question},
  Macmillan and Co.,
  London, 1866
}

\Lit{1936kz}{
  Konrad Zuse:
  \rtit{Die Rechenmaschine des Ingenieurs},
  Konrad Zuse Internet Archive,
  1936
}

\Lit{1945jn}{
  John von Neumann:
  \rtit{First Draft of a Report on the EDVAC},
  Moore School of Electrical Engineering, University of Pennsylvania,
  June 30, 1945
}

\Lit{1959ed}{
  Edsger W. D{\ij}kstra:
  \rtit{A Note on Two Problems in Connexion with Graphs},
  Numerische Mathematik~1, 1959, p.~269--271
}

\Lit{1965gm}{
  Gordon E. Moore:
  \rtit{Cramming more components onto integrated circuits},
  Electronics Magazine (Volume 38, Number 8),
  April 19, 1965
}

\Lit{1970cm}{
  Charles H. Moore:
  \rtit{Programming a Problem-Oriented-Language},
  June 1970
}

\Lit{1970jl}{
  Jan Łukasiewicz:
  \rtit{Selected works},
  Ed. by L. Borkowski,
  Amsterdam, London, North-Holland Publ. Comp., 1970
}

\Lit{1972bd}{
  W. J. Bouknight, Stewart A. Denenberg, David E. McIntyre,
  J. M. Randall, Amed H. Sameh, Daniel L. Slotnick:
  \rtit{The Illiac IV System},
  Proceedings of the IEEE Vol.~60, No.~4, April~1972
}

\Lit{1972mf}{
  Michael J. Flynn:
  \rtit{Some Computer Organizations and Their Effectiveness},
  IEEE Transactions on Computers, vol.~C-21, no.~9,
  September~1972,
  p.~948--960
}

\Lit{1973br}{
  Brian Randell (Ed.):
  \rtit{The Origins of Digital Computers},
  Springer, 1973
}

\Lit{1974as}{
  Alan C. Shaw:
  \rtit{The Logical Design of Operating Systems},
  Prentice Hall,
  1974
}

\Lit{1975ed}{
  Edsger W. D{\ij}kstra:
  \rtit{Essays on the nature and role of mathematical elegance},
  EWD619,
  Nuenen, The Netherlands,
  1975
}

\Lit{1975gm}{
  Gordon E. Moore:
  \rtit{Progress in digital integrated electronics},
  International Electron Devices Meeting, IEEE, 1975, p.~11--13,
  1--3~Dec. 1975
}

\Lit{1977nw}{
  Niklaus Wirth:
  \rtit{Compilerbau},
  Teubner, Stuttgart,
  1977
}

\Lit{1978ch}{
  C. A. R. Hoare:
  \rtit{Communicating sequential processes},
  The \break Queen's University, Belfast, Northern Ireland,
  Commun. ACM~21, \break 8 (Aug.~1978), p.~666--677
}

\Lit{1978kr}{
  Brian W. Kernighan, Dennis M. Ritchie:
  \rtit{The C Programming Language (1st ed.)},
  Prentice Hall,
  Englewood Cliffs,
  February 1978
}

\Lit{1979yc}{
  Edward Yourdon, Larry L. Constantine:
  \rtit{Structured Design},
  Prentice Hall,
  1979
}

\Lit{1980jp}{
  J. Postel:
  \rtit{Internet Protocol},
  Request For Comments RFC~760,
  DOD Standard, January~1980,
  and \newline
  J. Postel:
  \rtit{User Datagram Protocol},
  Request For Comments RFC~768,
  ISI, 1980
}

\Lit{1981jp}{
  J. Postel (Ed.):
  \rtit{Transmission Control Protocol},
  Request For Comments RFC~793,
  DARPA Internet Program, September~1981
}

\Lit{1981ps}{
  David A. Patterson, Carlo H. Séquin:
  \rtit{RISC I: A Reduced Instruction Set VLSI Computer},
  University of California, Berkeley,
  1981
}

\Lit{1983jf}{
  Joseph A. Fisher:
  \rtit{Very Long Instruction Word Architectures and the ELI-512},
  Yale University, New Haven,
  1983
}

\Lit{1985dh}{
  W. Daniel Hillis:
  \rtit{The Connection Machine},
  MIT Press, Cambridge, 1985
}

\Lit{1987bk}{
  W. Bibel, F. Kurfeß, K. Aspetsberger, P. Hintenaus, J. Schumann:
  \rtit{Parallel Inference Machines},
  in \rtit{Future Parallel Computers},
  Springer,
  1987,
  p.~185--226
}

\Lit{1987bv}{
  F. Baiardi, M. Vanneschi:
  \rtit{Parallelism Issues in Multi-Style Computers},
  in \rtit{Future Parallel Computers},
  Springer,
  1987,
  p.~1--34
}

\Lit{1987ms}{
  David May, Roger Shepherd, Catherine Keane:
  \rtit{Communicating Process Architecture: Transputers and Occam},
  in \rtit{Future Parallel Computers},
  Springer,
  1987,
  p.~35--81,
}

\Lit{1987pu}{
  David Peleg, Eli Upfal:
  \rtit{Efficient message passing using succinct routing tables},
  IBM Research Division,
  1987
}

\Lit{1987tr}{
  Philip C. Treleaven, Apostolos N. Refenes, Kenneth J. Lees, Stephen C. McCabe:
  \rtit{Computer Architectures for Artificial Intelligence},
  in \rtit{Future Parallel Computers},
  Springer,
  1987,
  p.~416--482
}

\Lit{1987wg}{
  Wolfgang K. Giloi:
  \rtit{Interconnection Networks for Massively Parallel Computer Systems},
  in \rtit{Future Parallel Computers},
  Springer,
  1987,
  p.~321--348
}

\Lit{1988gr}{
  Alan Gibbons, Wojciech Rytter:
  \rtit{Efficient Parallel Algorithms},
  Cambridge University Press,
  1988
}

\Lit{1988hj}{
  R. W. Hockney, C. R. Jesshope:
  \rtit{Parallel Computers 2},
  Second Edition, 1988
}

\Lit{1989dm}{
  David May:
  \rtit{The influence of VLSI technology on computer architecture},
  in \rtit{Scientific Applications of Multiprocessors},
  Prentice Hall,
  1989,
  p.~21--36
}

\Lit{1989eh}{
  Eds. R. J. Elliott, C. A. R. Hoare:
  \rtit{Scientific Applications of Multiprocessors},
  Prentice Hall,
  1989
}

\Lit{1989lv}{
  L. G. Valiant:
  \rtit{Optimally universal parallel computers},
  in \rtit{Scientific Applications of Multiprocessors},
  Prentice Hall,
  1989,
  p.~17--20
}

\Lit{1990lt}{
  \rtit{Digitale Fernsprechvermittlungstechnik -- Band 2: Das System EWSD},
  Bremen, L.T.U.-Vertriebsgesellschaft,
  1990
}

\Lit{1990tr}{
  Andrew S. Tanenbaum, Robbert van Renesse, Hans van Staveren, Gregory J. Sharp
  Sape J. Mullender, Jack Jansen, Guido van Rossum:
  \rtit{Experiences with the Amoeba Distributed Operating System},
  Dept. of Mathematics and Computer Science, Vr{\ij}e Universiteit, and
  Centrum voor Wiskunde en Informatica, Amsterdam,
  1990
}

\Lit{1991pp}{
  Dave Presotto, Rob Pike, Ken Thompson, Howard Trickey:
  \rtit{Plan 9, A Distributed System},
  in \rtit{Proceedings of the Spring 1991 EurOpen Conference},
  1991,
  p.~43--50
}

\Lit{1991ps}{
  Perihelion Software Ltd:
  \rtit{The Helios Operating System},
  May~1991
}
\Lit{1991tw}{
  Arthur Trew and Greg Wilson (Eds.):
  \rtit{Past. Present, Parallel:
  A Survey of Available Parallel Computing Systems},
  Springer,
  1991
}

\Lit{1992bf}{
  Stephen D. Brown, Robert J. Francis, Jonathan Rose, Zvonko G. Vranesic:
  \rtit{Field-Programmable Gate Arrays},
  Springer,
  1992
}

\Lit{1992in}{
  \rtit{i860 Microprocessor Family Programmer's Reference Manual},
  Intel, 1992
}

\Lit{1992ks}{
  K. Sollins:
  \rtit{The TFTP Protocol (Revision 2)},
  Request For Comments RFC~1350,
  Network Working Group, MIT, 1992
}

\Lit{1994os}{
  Oskar Schirmer:
  \rtit{Verlustfreie Datenkompression -- Überblick und Kategorisierung},
  Technische Universität Berlin,
  25.~November 1994
}

\Lit{1996ed}{
  Edsger W. D{\ij}kstra:
  \rtit{The next fifty years},
  EWD1243a,
  Austin, Texas,
  1996
}

\Lit{1996hv}{
  Horst Völz:
  \rtit{Informationsspeicher: Grundlagen -- Funktionen -- Geräte}
  expert-Verlag, Linde, 1996
}

\Lit{1998jb}{
  John E. Bjorkholm:
  \rtit{EUV Lithography -- The Successor to Optical Lithography?},
  Intel Technology Journal Q3'98,
  1998,
  p.~1--8
}

\Lit{2000rp}{
  Rob Pike:
  \rtit{Systems Software Research is Irrelevant},
  Bell Labs, Lucent Technologies,
  Feb~21, 2000
}

\Lit{2001ed}{
  Edsger W. D{\ij}kstra:
  \rtit{My recollections of operating system design},
  EWD1303,
  Austin, 2001
}

\Lit{2001zi}{
  \rtit{Z80 Family CPU Peripherals -- User Manual}
  (UM008101-0601),
  Zilog Inc.,
  2001
}

\Lit{2002cl}{
  Changhun Lee:
  \rtit{Distributed Shared Memory},
  Proceedings on the 15th CISL Winter Workshop, Kushu, Japan,
  February 2002
}

\Lit{2003kj}{
  \r{A}ge Kvalnes, Dag Johansen, Audun Arnesen, Robbert van Renesse:
  \rtit{Vortex: an event-driven multiprocessor operating system
  supporting performance isolation},
  June 13, 2003
}

\Lit{2004fm}{
  James Fung, Steve Mann:
  \rtit{Using Multiple Graphics Cards as a General Purpose Parallel Computer:
  Applications to Computer Vision},
  University of Toronto, 2004
}

\Lit{2004lf}{
  Laurie J. Flynn:
  \rtit{Intel Halts Development of 2 New Microprocessors},
  The New York Times,
  May~8, 2004
}

\Lit{2005cr}{
  Thomas Chen, Ram Raghavan, Jason Dale, Eiji Iwata:
  \rtit{Cell Broadband Engine Architecture and its first implementation},
  IBM,
  November~29, 2005
}

\Lit{2006dd}{
  Marc Daumas, Guillaume Da Graça, David Defour:
  \rtit{Caractéristi\-ques arithmétiques des processeurs graphiques},
  Université de Perpignan et Université de Montpellier~2,
  2006
}

\Lit{2006el}{
  Edward A. Lee:
  \rtit{The Problem with Threads},
  EECS Department, University of California, Berkeley,
  2006, January~10
}

\Lit{2006pf}{
  Paola Festa:
  \rtit{Shortest path algorithms},
  in \rtit{Handbook of Optimization in Telecommunications},
  Kluwer, Dordrecht,
  2006,
  p.~185--210
}

\Lit{2006th}{
  Tom R. Halfhill:
  \rtit{Ambric's New Parallel Processor},
  Microprocessor Report,
  October~10, 2006
}

\Lit{2007dm1}{
  David May:
  \rtit{Commodity High Performance Computing},
  Bristol, March 2007
}

\Lit{2007dm2}{
  David May:
  \rtit{Communicating Process Architecture for Multicores},
  CPA, Surrey, July~10, 2007
}

\Lit{2007hh}{
  David Money Harris, Sarah L. Harris:
  \rtit{Digital Design and Computer Architecture},
  Morgan Kaufmann,
  2007
}

\Lit{2007hv}{
  Horst Völz:
  \rtit{Handbuch der Speicherung von Information, Band 3:
  Geschichte und Zukunft elektronischer Medien},
  Shaker Verlag,
  Aachen,
  2007
}

\Lit{2008bc}{
  Silas Boyd-Wickizer, Haibo Chen, Rong Chen, Yandong Mao, Frans
  Kaashoek, Robert Morris, Aleksey Pesterev, Lex Stein, Ming Wu, Yuehua
  Dai, Yang Zhang, Zheng Zhang:
  \rtit{Corey: An Operating System for Many Cores},
  Proceedings of the 8th USENIX Symposium on Operating
  Systems Design and Implementation (OSDI '08), San Diego, California,
  December 2008, p.~43--57.
}

\Lit{2008es}{
  ECSS Secretariat:
  \rtit{SpaceWire -- Links, nodes, routers and networks},
  ECSS-E-ST-50-12C,
  ESA--ESTEC,
  31~July 2008
}

\Lit{2008gh}{
  Gardner Hendrie:
  \rtit{Oral History of Jean Bartik},
  Oaklyn, New Jersey,
  Recorded July 1, 2008
}

\Lit{2009bb}{
  Andrew Baumann, Paul Barham, Pierre-Evariste Dagand, Tim Harris, Rebecca Isaacs,
  Simon Peter, Timothy Roscoe, Adrian Schüp\-bach, Akhilesh Singhania:
  \rtit{The Multikernel: A new OS architecture for scalable multicore systems},
  SOSP'09, October~11--14, 2009
}

\Lit{2009mf}{
  Muddassar Farooq:
  \rtit{Bee-Inspired Protocol Engineering},
  Springer,
  2009
}

\Lit{2009mm}{
  David May, Henk Muller:
  \rtit{XCORE XS1 Architecture Tutorial,
  compact resources and instruction set description},
  Version 1.1,
  XMOS Ltd., 2009/6/22
}

\Lit{2009nh}{
  Edmund B. Nightingale, Orion Hodson, Ross McIlroy, Chris Hawblitzel, Galen Hunt:
  \rtit{Helios: Heterogeneous Multiprocessing with Satellite Kernels},
  SOSP'09, October~11--14, 2009
}

\Lit{2009sg}{
  Abraham Silberschatz, Peter Baer Galvin, Greg Gagne:
  \rtit{Operating System Concepts},
  8th Edition,
  John Wiley \& Sons,
  2009
}

\Lit{2009ti}{
  \rtit{OMAP-L138 C6000 DSP+ARM Processor},
  Texas Instruments,
  June 2009
}

\Lit{2009ts}{
  Thomas Sterling:
  \rtit{The Biggest Need: A New Model of Computation},
  Louisiana State University, March~30, 2009
}

\Lit{2009uc}{
  \rtit{The Unicode Standard, Version 5.2.0},
  The Unicode Consortium,
  Mountain View, CA, 2009
}

\Lit{2010dm}{
  David May:
  \rtit{XMOS Architecture, XC Language},
  XMOS Ltd.,
  London, April 2010
}

\Lit{2011cs}{
  \rtit{IBM's Server Processors: The RS64 and the POWER},
  The CPU Shack Museum,
  January~24th, 2011
}

\Lit{2011ga}{
  \rtit{F18A Technology Reference -- Product Data Book},
  GreenArrays, Inc.,
  12~April 2011
}

\Lit{2011hh}{
  James Hanlon, Simon J. Hollis:
  \rtit{Fast Distributed Process Creation with the XMOS XS1 Architecture},
  Communicating Process Architectures 2011, 33,
  2011,
  p.~195--207
}

\Lit{2011hp}{
  John L. Hennessy, David A. Patterson:
  \rtit{Computer Architecture: A Quantitative Approach},
  5th Edition,
  Morgan Kaufmann, 2011
}

\Lit{2011jm}{
  Jeff Martin:
  \rtit{Propeller Manual, Version 1.2},
  Parallax Inc.,
  2011
}

\Lit{2011pm}{
  Paul E. McKenney:
  \rtit{Is Parallel Programming Hard, And, If So, What Can You Do About It?},
  Linux Technology Center, IBM Beaverton,
  February~12, 2011
}

\Lit{2012ai}{
  \rtit{Epiphany Architecture Reference (G3)},
  Adaptevy Inc.,
  2012
}

\Lit{2012hm}{
  Thang Viet Huynh, Manfred Mücke, Wilfried N. Gansterer:
  \rtit{Evaluation of the Stretch S6 Hybrid Reconfigurable Embedded
  CPU Architecture for Power-Efficient Scientific Computing},
  Procedia Computer Science 9 (2012), p.~196--205
}

\Lit{2012sb}{
  Oskar Schirmer, Peter Börner:
  \rtit{XMOS -- der Echtzeit-Prozessor},
  Elektronik {\it embedded},
  Weka Fachmedien GmbH,
  Oktober 2012
}

\Lit{2013cs}{
  Chris Simmonds:
  \rtit{A timeline for embedded Linux},
  2net Ltd.,
  24th October 2013
}

\Lit{2013mf}{
  Mohamed Mahmoud Mohamed Farag:
  \rtit{ASIC Design Of The OpenSPARC T1 Processor Core},
  Faculty of Engineering, Cairo University, Giza, Egypt,
  2013
}

\Lit{2013os}{
  Oskar Schirmer:
  \rtit{Using Virtual Addresses with Communication Channels},
  2013
}

\Lit{2013tc}{
  \rtit{Tile Processor Architecure Overview for the TILEPro Series},
  Tilera Corporation,
  February 2013
}

\Lit{2014bg}{
  Muharrem Bayraktar, Fred A. van Goor, Klaus J. Boller, Fred B{\ij}kerk:
  \rtit{Spectral purification and infrared light recycling
  in extreme ultraviolet lithography sources},
  Optics Express, Vol.~22, No.~7,
  7~April 2014
}

\Lit{2014yb}{
  Xiangyao Yu, George Bezerra, Andrew Pavlo, Srinivas Devadas,
  Michael Stonebraker:
  \rtit{Staring into the Abyss: An Evaluation of
  Concurrency Control with One Thousand Cores},
  Proceedings of the VLDB Endowment, Volume~8, No.~3,
  Kohala Coast, Hawaii,
  November 2014
}

\Lit{2015jb}{
  James Bowman:
  \rtit{J1a SwapForth Reference},
  Excamera Labs, Pescadero, California USA,
  2015
}

\Lit{2015mv}{
  S. Mittal, J. Vetter:
  \rtit{A Survey of CPU-GPU Heterogeneous Computing Techniques},
  ACM Computing Surveys 47(4) p.~1--35,
  2015
}

\Lit{2016bb}{
  David Barron, Alexandre Bicas Caldeira, Volker Haug:
  \rtit{IBM Power System S821LC: Technical Overview and Introduction},
  IBM Corp.,
  December 2016
}

\Lit{2016ki}{
  \rtit{MPPA2(R)-256 Bostan Processor},
  Kalray Inc.,
  2016
}

\Lit{2016os1}{
  Oskar Schirmer:
  \rtit{NOP -- A Simple Experimental Processor for Parallel Deployment},
  Göttingen, 2016
}

\Lit{2016os2}{
  Oskar Schirmer:
  \rtit{GuStL -- An Experimental Guarded States Language},
  Göttingen, 2016
}

\Lit{2017xi}{
  \rtit{7 Series FPGAs Data Sheet: Overview},
  Xilinx, Inc.,
  DS180 (v2.4) March~28, 2017
}

\Lit{2018si}{
  Nikolay A. Simakov, Martins D. Innus, Matthew D. Jones, Joseph P. White,
  Steven M. Gallo, Robert L. DeLeon, Thomas R. Furlani:
  \rtit{Effect of Meltdown and Spectre Patches on the Performance of HPC Applications},
  January 2018
}

\end{list}

\end{document}